\documentclass[fleqn,usenatbib]{mnras}
\usepackage{subcaption}
\usepackage{graphicx}
\usepackage{array, makecell}

\usepackage[T1]{fontenc}
\usepackage{ae,aecompl}
\usepackage{tabularx}

\usepackage{graphicx}	
\usepackage{amsmath}	
\usepackage{amssymb}	
\usepackage{newtxtext,newtxmath}

\title[Pop III X-ray Binaries]{Population III X-ray Binaries and their Impact on the Early Universe}

\author[N. S. Sartorio et al.]{
Nina S. Sartorio,$^{1,2}$\thanks{E-mail: sartorio.nina@ast.cam.ac.uk}
A. Fialkov,$^{1}$
T. Hartwig,$^{4,5,6}$
G. M. Mirouh, $^{3}$
R. G. Izzard, $^{3}$
M. Magg,$^{7}$
\newauthor
R. S. Klessen,$^{7,8}$
S. C. O. Glover,$^{7}$
L. Chen.$^{7}$
Y. Tarumi,$^{4}$ and
D. D. Hendriks  $^{3}$
\\
$^{1}$Institute of Astronomy, University of Cambridge, Madingley Road, Cambridge CB3 0HA, UK\\
$^{2}$ Sterrenkundig Observatorium, Universiteit Gent, Krijgslaan 281 S9, 9000, Gent, Belgium\\
$^{3}$ Astrophysics Research Group, Faculty of Engineering and Physics, University of Surrey, Guildford GU2 7XH, UK\\
$^{4}${Department of Physics, School of Science, The University of Tokyo, Bunkyo, Tokyo 113-0033, Japan}\\
$^{5}${Institute for Physics of Intelligence, School of Science, The University of Tokyo, Bunkyo, Tokyo 113-0033, Japan}\\
$^{6}$ Kavli Institute for the Physics and Mathematics of the Universe (WPI), The University of Tokyo Institutes for Advanced Study, Japan\\
$^{7}$ Universität Heidelberg, Zentrum für Astronomie, Institut für theoretische Astrophysik, Albert-Ueberle-Str. 2, 69120 Heidelberg, Germany \\
$^{8}$ Universit\"{a}t Heidelberg, Interdisziplin\"{a}res Zentrum f\"{u}r Wissenschaftliches Rechnen, Im Neuenheimer Feld 205, 69120 Heidelberg, Germany}

\date{Accepted 2023 February 28. Received 2023 February 28; in original form 2022 December 07}

\pubyear{2015}
\usepackage[dvipsnames]{xcolor}

\begin{document}
\label{firstpage}
\pagerange{\pageref{firstpage}--\pageref{lastpage}}
\maketitle

\begin{abstract}
The first population of X-ray binaries (XRBs) is expected to affect the thermal and ionization states of the gas in the early Universe. Although these X-ray sources are predicted to have important implications for high-redshift observable signals,  such as the hydrogen 21-cm signal from cosmic dawn and the cosmic X-ray background, their properties are poorly explored, leaving theoretical models largely uninformed. In this paper we model a population of X-ray binaries arising from zero metallicity stars. We explore how their properties depend on the adopted initial mass function (IMF) of primordial stars, finding a strong effect on their number and X-ray production efficiency. We also present scaling relations between XRBs and their X-ray emission with the local star formation rate, which can be used in sub-grid models in numerical simulations to improve the X-ray feedback prescriptions. Specifically, we find that the uniformity and strength of the X-ray feedback in the intergalactic medium is strongly dependant on the IMF. Bottom-heavy IMFs result in a smoother distribution of XRBs, but have a luminosity orders of magnitude lower than more top-heavy IMFs. Top-heavy IMFs lead to more spatially uneven, albeit strong, X-ray emission. An intermediate IMF has a strong X-ray feedback while sustaining an even emission across the intergalactic medium. These differences in X-ray feedback could be probed in the future with measurements of the  cosmic dawn 21-cm line of neutral hydrogen, which offers us a new way of constraining population III IMF.
\end{abstract}




\section{Introduction}

X-rays are thought to be one of the main drivers of the evolution of the baryonic component in the early Universe thanks to their impact on early star formation \citep{Park_2021}  as well as heating and ionizing the intergalactic medium \citep[IGM, e.g., ][]{Fialkov_2014, Pacucci:2014, Das_2017, Madau_2017, Ross_2017}. Most of the X-ray emission today is produced by active-galactic nuclei \citep[AGN: ][]{2013_Fragos, 2022Kovlakas}. However, at redshifts $z> 4$, the number of AGN quickly decreases due to the age of the Universe being too short to allow for the formation of supermassive black holes in most galaxies \citep{Treister_2012, Vito_2018}. As a result, at the beginning of the Epoch of Reionization (EoR, $z\sim6-15$), X-ray binaries (XRBs) are thought to dominate the X-ray photon budget \citep{2013_Fragos}.

At these times, the IGM was predominantly neutral and shielded from most forms of radiation produced by  first stars and stellar remnants. Soft X-rays ($0.1 - 2$ keV) are one of the few frequency bands that are expected to have a significant impact on the high-redshift IGM \citep{2001Venkatesan,2003_Glover,Fragos_2013a, Fialkov_2014, Pacucci:2014, Eide_2018} as these photons are able to both escape from their host dark matter halos and have a mean free path smaller than the size of the Universe, thus injecting their energy into the gas. 

X-ray photons lead to significant changes in the thermal and ionization states of the IGM, thus affecting the observable  21-cm signal of neutral hydrogen, which is one of the most promising probes of this era \citep[e.g.,][]{Fialkov_2014, Pacucci:2014, Das_2017, Fialkov_2017, Eide_2018, EW:2018, Ma:2018, Munoz:2021, 2022Kovlakas}. Therefore, radio telescopes, such as interferometers HERA \citep{DeBoer2017}, PRIZM \citep{PRIZM} LOFAR \citep{Gehlot:2020}, LEDA \citep{Garsden:2021}, MWA \citep{McKinley2018}, NenuFAR \citep{Mertens:2021}  and the upcoming SKA \citep{Koopmans2015} and radiometers including EDGES \citep{Bowman:2018}, SARAS \citep{Singh:2021} and REACH \citep{2022REACH}, have the potential to constrain properties of the first population of X-ray binaries.  Recent upper limits on the 21-cm signal established by  some of these telescopes,  although still weak, disfavour cold IGM at the observed redshifts indicating that some amount of X-ray emission was produced prior to the completion of reionization \citep{Singh:2018, Monsalve2019, LOFAR1, LOFAR2, LOFAR3, MWA_fX, HERAtheory, 2022bevins}.

In contrast, harder X-rays, with energies exceeding a few keV, are never absorbed due to their large mean free paths \citep{2007pritchard, Fialkov_2014}, and instead contribute to the cosmic X-ray background \cite[CXB, e.g., ][]{Christian:2013, Fialkov_2017, EW:2018, Ma:2018}.  A fraction of the observed CXB, detected with telescopes such as XMM-Newton, Chandra and Swift \citep{Lumb_2002, De_luca2004, Hickox_2006, Moretti_2012}, remains unresolved and must include any high-redshift X-ray contribution. However, the magnitude of this contribution is poorly known  and depends on the unconstrained astrophysics of the high-redshift Universe \citep[e.g., ][]{Cappelluti:2012,2013_Fragos, Fialkov_2017, EW:2018, Ma:2018}. 

In addition to heating and ionizing the IGM, X-rays may also have played a role in early star formation \citep[][Klessen \& Glover 2022]{2012_Yoshida,2013_bromm, Jeon_2012}. Gas from which the first stars were formed (Population III or Pop III stars) was almost completely devoid of metals. Hence, in order to condense and collapse into stars, this gas depended on cooling by molecular hydrogen ($\text{H}_2$). The abundance of $\text{H}_2$ itself was conditioned by the intensity of  UV emission  at the Lyman and Werner absorption bands of molecular hydrogen ($11.2-13.6$ eV). This Lyman-Werner radiation was able to dissociate $\text{H}_2$ and, although it probably did not completely shut down star formation, it may have considerably slowed it down \citep{2020skinner,Wise_2007, Safranek-Shrader_2012, Schauer_2017, Schauer_2020}. In contrast, X-rays have the opposite effect on the $\text{H}_2$ abundance. As X-rays ionize hydrogen atoms, the number of free electrons increases. These electrons act as catalysts for the formation of $\text{H}_2$, and thus, increase the ability of gas to cool \citep{2001oh,2003_Glover, Machacek2003}. This extra cooling from increased $\text{H}_2$ abundance typically overcomes the effect of X-ray heating whenever densities exceed a few 10 cm$^{-3}$, and thus aids star formation in dense regions \citep{Hummel_2015, Park_2021}. How large a role the X-ray background plays in regulating the formation of Pop III stars is still debated \citep{Ricotti_2016, Park_2021} .

Given the importance of X-rays in the early universe, it is crucial that we model their emission as accurately as possible. In this paper we focus on the population of XRBs resulting from metal-free Pop III stars. In order to bracket the uncertainty in the yet unknown initial mass function (IMF) of the first stellar population \citep{Chen_2020}, we consider three very distinct IMFs. This allows us, for the first time, to predict the IMF dependency of the luminosity, spectral energy distribution (SED) and CXB created by the population of first XRBs at redshifts $z>10$. Similar work has been done by \citet{Fragos_2013a},  but focusing on Pop II (metal-poor) stars and using a bottom-heavy IMF similar to that adopted at solar metallicity \citep{Kroupa2001,Kroupa_2003}. 

This paper is organised as follows. In Section \ref{sec: lm hm} we start by presenting a discussion on XRBs and their current categorisation. We explain how the large typical mass and zero metallicity of the first stars lead to a population of XRBs distinct from the observed one.
In Section \ref{sec: Methods} we outline the methodology.  In Section \ref{sec: params that lead to XRBs} we analyse which binary systems, out of the broadly defined  initial sample, become XRBs. In Section \ref{sec:results}, we lay out our main results. We discuss how the X-ray feedback from each IMF differs and the implications this difference has for the X-ray feedback on the IGM. We also provide simple scaling relations between the number of XRBs, the X-ray luminosity and the star formation rate. 
The data presented can be used in semi-analytical modelling and as  sub-grid prescriptions in simulations. Finally, the conclusions are found in Section \ref{Sec:conclusions}.

\section{Population III XRBs: How are they different? }
\label{sec: lm hm}
XRBs are binary systems in which the primary object is a black hole or a neutron star that accretes material from a non-compact companion, often referred to as the secondary object, or simply the secondary. As material is transferred from the secondary to the primary, it heats up and emits copious amounts of X-rays. 

These binaries are often subdivided into low and high mass XRBs  (LMXBs/HMXBs) depending on the mass of the  companion  ($ M_\star \sim  {\rm M}_\odot$ for the former and $M_\star > 10 \, {\rm M}_\odot$ for the latter)\footnote{Sometimes the secondaries/companions in XRBs which have masses $1{\rm M}_\odot  < M_\star < 10 {\rm M}_\odot $ are referred to as intermediate mass X-ray binaries (IMXBs). However, this classification is less common. }. 
In this paper, for ease of comparison with previous works, we adopt the definition of \citet{Fragos_2013a} in which $3\,{\rm M}_\odot$ acts as a dividing mass between LMXBs and HMXBs.

With regards to present-day observed XRBs, these two categories of XRBs usually correlate with distinct methods of mass transfer between the primary and the secondary, i.e. the compact object and its companion. In LMXBs the secondary usually evolves off the main sequence, fills its Roche lobe and begins to transfer mass to the primary via Roche lobe overflow (RLOF). As mass is transferred from the companion through the first Lagrange point towards the compact object, an accretion disc forms around the primary and becomes the main source of X-rays. 

HMXBs sometimes also transfers material by RLOF, however the majority of observed HMXBs seem to rely on other accretion modes \citep{Coleiro_2013, Chaty_2013}. Most detected HMXBs are BeXRBs, that is, XRBs in which the secondary is a Be star which is a B-type star that spins with velocities between 0.5 to 0.9 of its critical velocity \citep{Belczynski_2009, Antoniou_2016}, whose rapid rotation around its axis ejects material which distributes itself in the shape of a disk around the star giving rise to a decretion disk \citep{Rappaport_1982}. In a BeXRB,  X-ray emission happens whenever the compact object crosses the decretion disc of the Be star thereby accreting some of its mass \citep{Brown_2019,Vinciguerra_2020}. The remainder of the HMXB population are Supergiant High-Mass XRBs. In these XRBs X-ray emission is powered by accretion of a strong wind, or by RLOF, with the latter being less common. 

High-redshift XRBs originating from Pop III stars could be substantially different from their present-day counterparts because stellar population properties depend on metallicity, and, thus, redshift. First, the binary fraction of Pop III stars is not known and we have to rely on simulations to make estimates. A number of studies have shown that, just like a large fraction of stars today \citep{2006lada, 2021Luo}, Pop III stars are expected to be born in multiples  \citep{Turk_2009,Greif_2011, Satcy_2013, Stacy_2014,Susa_2019}. In particular, \citet{Satcy_2013} predict a binary fraction of 36\%, that is a given star has a 50\% probability of being in a binary, which we adopt in this study. Though the binary fraction is poorly constrained, we expect our results to depend linearly on its value such that the prescriptions presented here can be easily corrected if a different fraction is adopted. In addition, evolution of a zero-metallicity Pop III star is expected to differ from that of a more metal-rich star. In the following we comment on a few aspects in which those differences may matter for the formation and properties of XRBs.

\subsection{Winds}
One of the most important differences between stellar populations are the line driven winds. At present day, massive stars lose a substantial fraction of their mass via winds during their lifetime. These winds are radiation-driven, with the wind acceleration relying on multiple spectral lines of metals that are able to absorb significant momentum from the incoming photons \citep{Castor_1975}.


In contrast, Pop III stars cannot drive winds effectively because they form from metal-free gas \citep{Puls_2008}. In this scenario, the only appreciable sources of opacity are the lines of singly ionized helium, Thomson scattering by free electrons and lines of any carbon, nitrogen and oxygen which the primordial star synthesises throughout its lifetime and that mix into the stellar atmosphere.  We find that as Pop III stars evolve, and their core temperatures increases, carbon is produced by the strongly temperature-sensitive triple-$\alpha$ reaction. This creates enough metals to drive a weak CNO wind, which is present in the binaries modelled in this work. This makes most Pop III stars poor candidates for strong wind production as even a CNO-driven wind is relatively weak and does not lead to considerable mass losses \citep{Kudritzki_2002,Krticka_2009}. Other types of wind, such as dust-driven wind which is an important mass loss mechanism during the AGB (Asymptotic Giant Branch) phase. also scale with metallicity and are, generally, much weaker for Pop III stars \citep{2008Wachter,Tashibu16, 2018takeru}. As mentioned previously, it is thought that a large number of present-day HMXBs are wind-fed. We, thus, expect that these binaries would represent a small fraction of Pop III XRBs. These winds, albeit weak, are included in our modelling (see Section \ref{subsec: binaryC}).


\subsection{Rotation}
Present-day BeXRBs, which form a large fraction of HMXBs rely on rapid stellar rotation which inversely correlates to metallicity \citep{2006Chiappini, 2020Amard}. As discussed above, stellar winds in Pop III stars are weaker, leading to less mass and angular momentum being lost. Thus, Pop III stars should rotate faster on average than stars today \citep{Ekstrom_2008,Choplin_2019}.  If that is indeed the case, it is possible the fraction of HMXBs represented by BeXRBs was larger in the past than at present. However, in our models do not consider decretion discs due to rapidly rotating stars and, thus, ignore the possibly large contribution of BeXRBs which we will explore in a follow up paper. 

\subsection{Initial mass function}
The IMF of Pop III stars is very uncertain and adopted prescriptions vary significantly between studies. Early works favoured a very top-heavy IMF, with masses in excess of 100 M$_\odot$ \citep{Omukai_2001,Abel_2002, Bromm_2002}. However, in the past decade, more studies have shown that fragmentation occurs more efficiently than previously anticipated in metal-free clouds and could lead to the formation of a larger number of stars with solar-like masses \citep{Stacy_2010, Greif_2011, Clark_2011, Hirano_2014, Susa_2019, Sharda_2019, 2020_Wollenberg}. Although some works report sub-solar Pop III masses \citep{Stacy_2014}, the non-detection of metal-free stars to date  suggests a minimum stellar mass of Pop III stars of at least 0.7 M$_\odot$ \citep{Hartwig_2015, Magg_2019, Rossi_2021}. The IMF determines the ratio of low to high mass stars as well as the ratio of LMXBs to HMXBs. To bracket this uncertainty, in this paper, we consider three different the primordial IMFs as discussed in Section \ref{subsec:catalogues} and summarised in Table \ref{table: IMFs used}.






\section{Method}
\label{sec: Methods}

We explore the dependence of X-ray emission of Pop III XRBs on the assumed IMF of the stars. We start by creating catalogues of initial primordial binary stars (Section \ref{subsec:catalogues}) for distinct IMFs. Each binary is assigned orbital parameters, an eccentricity, $e$, and a period, $P$, (Section \ref{sec: OrbParams}) assuming these are distributed similarly to the orbital parameters of present-day binary systems. It is unknown how these orbital parameters are expected to differ in high-redshift systems, however, they are probably related to the mass of the stars in the binary. Thus, where possible we use a mass-dependent method to assign these parameters.  We then run each catalogue (Section  \ref{subsec: binaryC}) through the population evolution code {\sc{binary\_c}} \citep{Izzard_2004, Izzard_2006} to which we added a new X-ray emission module described in Section \ref{subsec: ModelEmission}. {\sc{binary\_c}} evolves each binary pair individually, tracking mass transfer and orbital parameter evolution over time. 
We consider, for the first time, zero-metallicity evolution and XRB formation. The evolution of the Pop III stars is based on an interpolation of metal-free stellar evolution models without rotation evolved by MESA \citep{2018_Paxton, 2019_Paxton}.\footnote{ {\sc{binary\_c}} uses MESA tables for now only for the main sequence evolution. For later evolutionary stages the code reverts to a binary stellar evolution scheme as in \citet{Hurley_2002}}


Whenever a binary system becomes an X-ray binary, i.e. a compact object accreting from a companion star, we compute the X-ray spectrum in 50 energy bins spanning frequencies from $10^{10}$ to $10^{23}$ Hz (corresponding to photon energies between $\sim 10^{-7}$ to $10^{6}$ keV) as well as how long each binary is an XRB. Finally, we compute local absorption of X-rays by hydrogen and helium within dark matter halos in which the XRBs are situated (Section \ref{sec:absorption}).


\subsection{Initial binary catalogues}
\label{subsec:catalogues}

We create catalogues of metal-free binaries at redshifts between $z=5$ and $z=30$. These catalogues are produced using the semi-analytical code {\sc{a-sloth}}  \citep[Ancient Stars and Local Observables by Tracing Haloes, ][]{Hartwig_2022, Magg2022} with an improved sub-grid model of stochastic metal mixing in the first galaxies \citep{Tarumi_2020}. This semi-analytical model follows a cosmologically representative sample of dark matter merger trees based on extended Press-Schechter theory \citep{Ishiyama_2016}. Based on the mass in each halo, a Pop III star formation rate is calculated. A necessary condition for Pop III star formation is that cooling by molecular hydrogen is sufficient to induce fragmentation, which requires that the halo mass is above the critical mass
\begin{equation}
    M_\mathrm{crit} = 3\times 10^6\,{\rm M}_\odot \left( \frac{1+z}{10} \right)^{-3/2},
\end{equation}
corresponding to a virial temperature $T_\mathrm{vir}=2200\,\mathrm{K}$ \citep{Hummel_2012,Glover_2013}\footnote{Although there is some uncertainty in the value and redshift evolution of $M_{\rm crit}$, this simple prescription agrees relatively well with the results of \citet{2019schauer} using an extremely high resolution cosmological simulation paired with a state-of-the-art primordial chemistry model (Klessen \& Glover (submitted)).}. Furthermore, in order to be labelled as a Pop III star, the gas in the halo must be pristine (i.e. the chemical composition of the gas is that resultant from Big Bang nucleosynthesis). Once these conditions are met, stars are sampled from a chosen IMF and assigned to a specific halo. The formation of Pop III stars is followed over time, including self-consistent chemical, radiative and mechanical feedback \footnote{Although {\sc{a-sloth}}  has self-consistent radiative feedback when creating the stellar catalogues it does not include an X-ray feedback. As such, the impact of IMF-dependant X-ray emission on star formation is not taken into account. We intend to implement a new X-ray prescription in {\sc{a-sloth}} according to the results presented here at a later date.} which considers the spatial position of the halos \citep[see][]{Tarumi_2020}. We follow a cosmologically representative comoving volume of 8 $(\rm{Mpc}/h)^3$ (where h is Hubble's constant in units of 100 km/s/Mpc) create our binary catalogues. At each timestep, a fraction of stars is selected to be newly-formed binary systems. We assume a binary fraction of 36\% \citep{Satcy_2013}. We pair the binaries randomly, such that there is no bias in the masses of the stars in the binary. 

Each catalogue includes the following information on the binary:

\begin{itemize} 
    \item the zero-age main-sequence (ZAMS) masses of the stars in the binary components,
    \item the redshift at which stars are formed and
    \item the stellar mass of the dark matter halo to which the binary system belongs. 
\end{itemize}
The redshift of each system in our binary catalogues is specified at the time when both stars in the binary, which we assume to form simultaneously, are on the ZAMS. Typically, there is a delay between the ZAMS and the redshift at which a system produces X-rays which itself depends on the binary parameters (stellar masses, periods and eccentricity). We add this delay experienced by each binary pair to their initial redshift, such that it is taken into account. 

Because the fate of the binary pair is dependent on the masses of the stars, we use {\sc{a-sloth}} to create three catalogues with distinct IMFs. Two of these IMFs represent extreme scenarios. The first is a bottom-heavy IMF with a Salpeter slope, referred as `Low-Mass'. In the other extreme we adopt a log-flat IMF, labelled as `High-Mass' containing only stars in excess of 10 ${\rm M}_\odot$.  Finally, we consider a more plausible,  `Fiducial' IMF which was calibrated to reproduce observables of the present-day Milky Way, such as the metallicity distribution function \citep{Tarumi_2020}. The properties of the adopted IMFs are summarised in Table \ref{table: IMFs used}. 

\begin{table}
\centering
\setcellgapes{3pt}\makegapedcells
\begin{tabular}{|c|c|c|c|} \hline
IMF Name &\makecell{\textbf{Low-Mass}\\(LM)} & \makecell{\textbf{Fiducial}\\(Fid)} & \makecell{\textbf{High-Mass}\\(HM)}\\ \hline
\makecell{Min. Mass \\ (${\rm M}_\odot$)}  &  0.8& 2 & 10 \\ \hline
\makecell{Max. Mass \\ (${\rm M}_\odot$)} &  250& 180 & 1000  \\ \hline
\makecell{ Slope \\ (dN/dM))} &\makecell{ -2.35 \\ (Salpeter)}  & -0.5 & \makecell{0 \\ (flat) } \\ \hline
\makecell{Num. of binaries}  & $37.5\times10^{6}$ & $1.7\times10^{6}$  & $0.3\times10^{6}$\\
\hline\end{tabular}
    \caption{The  three IMFs used in this work: a bottom-heavy (LM, left column), an intermediate (Fid, middle column) and a top-heavy IMF (HM, right column). For each IMF we list the minimum and maximum masses of stars adopted, as well as the slope of the IMF. We also show the number of binaries present in the catalogue for each IMF.}
    \label{table: IMFs used}
\end{table}

Note the stellar mass in each star forming halo at any time is independent of the chosen IMF. Thus, the more top-heavy the IMF adopted the fewer stars and binaries are sampled per halo. This leads to two orders of magnitude variation in the number of binaries between catalogues as shown in the last line of Table \ref{table: IMFs used}.

\subsection{Initial orbital parameters}
\label{sec: OrbParams}

Before we evolve each binary system we assign the binary an eccentricity, $e$, and a period, $P$, which together also determine the mean separation between the two stars. During the subsequent evolution in {\sc{binary\_c}} these orbital parameters evolve according to the interaction between the two stars and any mass or angular momentum loss.


\subsubsection{Eccentricity}

As we have no stringent constraints on the properties of Pop III binaries, we adopt a  physically-motivated distribution of eccentricities.
We assume that the initial eccentricities of our binaries follow a thermal eccentricity distribution:
\begin{equation}
    f(e) = 2e \, de.
\end{equation}
Here all the values of $e^2$ have the same likelihood, which implies that, on average, there are more binary systems with high eccentricities than with low. This description comes from the expectation that if a population of binaries undergoes enough dynamical encounters then it would eventually achieve energy equipartition, and, thus, the energy should follow a Boltzmann distribution \citep{Jeans_1919}. This prescription  has been used in a number of binary studies including those involving Pop III stars \citep[e.g.][]{Hartwig_2016}.

\subsubsection{Periods}
We use a combination of period distributions based on current observations. Binary systems with at least one OB star (i.e. stars with masses greater than $2 {\rm M}_\odot$) are best described by the period distribution of \citet{Sana_2012}:
\begin{equation}
    f_{\rm Sana}\left(\log P\right) \propto \left(\log \left(P/{\rm days}\right)\right)^{-0.55},
\end{equation}
which reflects the periods of the most massive binaries we know.

Less massive binaries (e.g. containing solar type stars) are observed to follow a different period distribution \citep{Kroupa_1995}:
\begin{equation}
    f_{\rm Kroupa}\left[\log P\right] = \frac{2.5(\log P -1)}{45+ \left(\log(P) -1\right)^2}.
\end{equation}

As a means to describe the entire population of binaries, we combine the two limits \citep[inspired by][]{Izzard_2018} and  obtain a mass dependent distribution 
\begin{equation}
    f_P = f_{\rm Kroupa}(1 - { M}/{ M}_{\rm max}) + f_{\rm Sana}({ M}/{ M}_{\rm max}),
\end{equation}
where M is the most massive star present in the binary.

We choose a period range from $0.15 < \log (P/{\rm days}) < 6.7 $, such that we cover the periods considered both by \citet{Kroupa_1995} and \citet{Sana_2012}.  Since these distributions were derived from observations of current massive binaries, which have  smaller masses than the most massive Pop III stars, we set $M_{\rm max} = 150 {\rm M}_\odot$ with larger stellar masses being simply sampled from $f_{\rm Sana}$. 



\subsection{Binary evolution prescriptions}
\label{subsec: binaryC}

With the masses and orbital parameters fixed as explained above, the binary pairs  are evolved using the population synthesis code {\sc{binary\_c}} \citep{Izzard_2004, Izzard_2006, Izzard_2009, Izzard+2018}, originally based on the Binary Star Evolution (BSE) code of \citet{Hurley_2002}. The code simulates the evolution of Pop III stars in each binary from the ZAMS until they become compact remnants.  

{\sc{binary\_c}} models the interaction between the stars in the binary such as mass transfer and tidal effects. Because stars in the most massive binaries are expected to interact via mass exchange at some point during their lifetimes \citep{Sana_2012}, this leads to a very distinct result from evolving the stars in isolation. 

{\sc{binary\_c}} has recently seen a number of modifications that allowed us to self-consistently evolve Pop III binaries. The updates include a new treatment of pair-instability supernovae, an improved stellar wind prescription \citep{Schneider_2018,Sander_2020} and stellar evolution at zero metallicity. Moreover, specifically for this work, we developed a new  {\sc{binary\_c}} module  that allows us to calculate  X-ray emission from XRBs. We present this module in Sec. \ref{subsec: ModelEmission}. By having the information of the binary parameters, both initially and at the time a binary emits X-rays, we analyse how each parameter affects the formation of XRBs ( Section \ref{sec: params that lead to XRBs}). 

\subsubsection{Winds in {\sc{binary\_c}}}

{\sc{binary\_c}} has many ways in which winds from massive stars can be incorporated. Here we use the same wind prescription as in  \citet{Schneider_2018} and \citet{Sander_2020} but at zero metallicity and hence with little radiatively-driven mass loss (i.e. on the main sequence). However, later phases of stellar evolution do have contribution of winds.

\subsubsection{End products of Pop III stars}

Because the luminosity and SED of an XRB strongly depends on the mass and type of its compact object it is important to  model the end products of binaries accurately. In {\sc{binary\_c}} we account for normal core-collapse supernovae as well as other possible scenarios such as pulsational pair-instability supernovae, pair-instability supernovae and photo-disintegrations. This is required since very massive Pop III stars have different evolutionary channels compared to present-day stars. Notably, stars around $\sim$140 ${\rm M}_\odot$ develop helium core masses in excess of 30 $\rm {\rm M}_\odot$. The following evolutionary stages of these stars are unstable and lead to the production of electron-positron pairs. The pairs form at the expense of thermal pressure support, leading to a rapid contraction of the core and associated increase in temperature and density. In turn, this triggers an explosive oxygen burning which unbinds the star and  leaves no compact remnant behind. Due to their believed higher masses and the low wind mass loss throughout their lifetimes, Pop III stars have been prime candidates to experience this so called “pair-instability” supernova  \citep[PISN,][]{Woosley_2007, Woosley_2010, Chen_2014} . 
A modelling of this phenomenon is of particular importance for this work as the absence of a remnant implies that a range of binaries where the most massive star is around $\sim 180-260~{\rm M}_\odot$ will never be able to yield an XRB \citep{Heger_2003, Farmer_2019}\footnotetext{It should be noted that there are uncertainties in the range of masses that lead to a PISN and that considerations such as rotation lead to different mass ranges that undergo PISN \citep{2020_Marchant, 2021Woosley}}. By reducing the range of masses, we reduce the number of potential black holes present at any time and furthermore we restrict the mass of black holes that do form.

For stars with less massive He-cores these nuclear flashes still occur but are not energetic enough to unbind the star leading instead to successive pulsations. In this case the core contracts, burns oxygen, expands and cools in a cyclic manner. These stars undergo pulsational pair-instability supernova (PPISN) and eventually die by a core collapse similar to other stars, but the mass loss due to the pulsations means the remaining black hole has a much smaller mass than one would naively expect from the initial mass of those stars. 

When the core is even more massive, the energy released by the oxygen
burning is mostly lost to neutrinos and to the photo-disintegration of the material in the core. Photo-disintegration is the process where energetic photons break up nuclei, which generally is an endothermic process when the nuclei are lighter than iron. This endothermic process removes the energy available for the explosion, preventing the reversal of the collapse. This leads to a core collapse event that is unable to successfully produce a supernova event, and the formation of a massive black hole \citep{Heger_2003,Yoon_2012, Habouzit_2016}.

For this work we model all three of these scenarios adopting the prescription from \citet{Farmer_2019}. In our modelling, for stars in isolation the initial masses in the ranges 110-180 ${\rm M}_\odot$, 180-260 ${\rm M}_\odot$ and $>260$  ${\rm M}_\odot$ would lead respectively to PPISN, PISN and photo-disintegration of the star. In practice, due to mass loss/gain during evolution, binaries modelled with {\sc{binary\_c}} that start in one of these mass ranges will not necessarily lead to the corresponding end product. 

\subsection{Model X-ray emission}
\label{subsec: ModelEmission}

XRBs produce X-ray photons via several processes including stellar emission and radiation from  accretion discs and jets.

The observed SEDs of XRBs seem to oscillate between having a peak at a few keV (soft/low energy X-rays) and a peak at a few hundred keV (hard/high energy X-rays). This  observed dichotomy of the XRB spectra motivates the definition of two distinct states, the {\it soft} and the {\it hard} states, of an X-ray binary. Theoretically, the existence of the two states can be explained by two distinct stable accretion flow structures: one dominated by an optically thick, but geometrically thin disc and another dominated by a hot, optically thin, geometrically thick corona. The former reproduces well the observed soft state SEDs while the latter can generate hard photons through thermal Comptonization of disc photons by the corona \citep{Zdziarski_2004}. 

In order to create a realistic spectrum for the binaries in this work, we generate an SED in steps, first deriving a disc spectrum consistent with accretion rates and masses of the accretors, and, secondly, Comptonising a fraction of the disc photons towards higher energy. Below we explain in detail how we accomplish these steps.

\subsubsection{Disc spectrum}

The XRBs we trace are binaries in which a star transfers material to its compact companion mainly via RLOF. In these systems the gas from the envelope of the secondary star overflows through the first Lagrange point towards the primary at radial velocities that are usually low enough for an accretion disc to form. This disc produces most of the soft X-ray emission and is the base of all radiation we will model for the XRB. We also have a number of stars that transfer mass via winds, for which we assume a disc can also form. 

We assume that the disc is a Shakura-Sunyaev type thin disc \citep{Shakura_1973} with a steady flow (constant accretion rate) and that the local angular velocity of the flow is Keplerian. It can be shown that, under these circumstances, the rate of viscous energy dissipation into heat is independent of the viscosity itself and that the disc energy dissipation per unit area per unit time, $\epsilon$, as a function of the disc radius, $r$, is given by:
\begin{equation}
    \epsilon(r) = \frac{3{\rm{G}M} \dot{ M}}{8 \pi r^3}\left[1 -\left(\frac{R_{\star}}{r}\right)^{1/2}\right],
\end{equation}
where G is the gravitational constant and $M$, $R_{\star}$ and $\dot{M} $ are the mass, radius and mass accretion rate of the accreting object \citep{Shakura_1973}. These parameters are computed by {\sc{binary\_C}} at the point at which binary is in the XRB phase. In the scenarios that we consider here, the radius of the compact object, $R_\star$, is typically much smaller than the innermost ring of the disc, such that the $R_\star/r$ term in the above equation is small and can be ignored.

We can model the outgoing spectrum with a spectrum of a multicolor disc \citep{Mitsuda_1984}, that is, a disc in which each annulus emits as a blackbody of a certain temperature, $T$, regulated by the conversion of kinetic into heat energy. 

The temperature at a given radius is given by the Stefan-Boltzmann law, $\epsilon(r) = \sigma T^4$, where $\sigma$ is the Stefan–Boltzmann constant,  which gives us the following expression for $T$:
\begin{equation}
 T(r) = \left(\frac{3{\rm{G}M} \dot{M} }{8 \pi r^3 \sigma}\right)^{1/4}.
\end{equation}

We can express the blackbody emission, $B$ (the power per unit area per frequency per solid angle), at a frequency $\nu$ of a given annulus of the disc at a radius r away from the accreting object as:

\begin{equation}
    B(\nu ,T(r))={\frac {2h\nu ^{3}}{c^                      {2}}}{\frac {1}{e^{h\nu/kT(r)}-1}}
\end{equation}
with $h$, $k$, $c$ being the Planck constant, the Boltzmann constant and the speed of light, respectively. 

The luminosity of all the annuli together is, thus, given by integrating the blackbody spectrum for each radius of the disc:
\begin{equation}
    L_\nu \propto \int_{R_{\text{min}}}^{R_{\text{max}}} 2\pi r B dr,
\end{equation}

We assume that the inner radius of the disc corresponds to the innermost stable circular orbit (ISCO), that is, $R_{\rm min} = 6\rm{GM}/c^2$, whereas the outer radius is set to correspond to 90\% of the Roche radius\footnote{Assuming the primary is the accretor and the secondary the donor, $R_{\rm Roche} = R_{\text{primary}}\left[(2M_{\text{primary}})/(M_{\text{secondary}})\right]^{1/3}$.} of the compact object, $R_{\rm max} = 0.9 R_{\rm Roche}$, the value suggested by observations \citep[e.g.][]{Paradijs1981}. The existence of such disc SEDs have been confirmed by detections in a number of luminous XRBs \citep{Davis_2006,Dunn_2011}.

\begin{figure}
    \centering
    \includegraphics[width=\columnwidth]{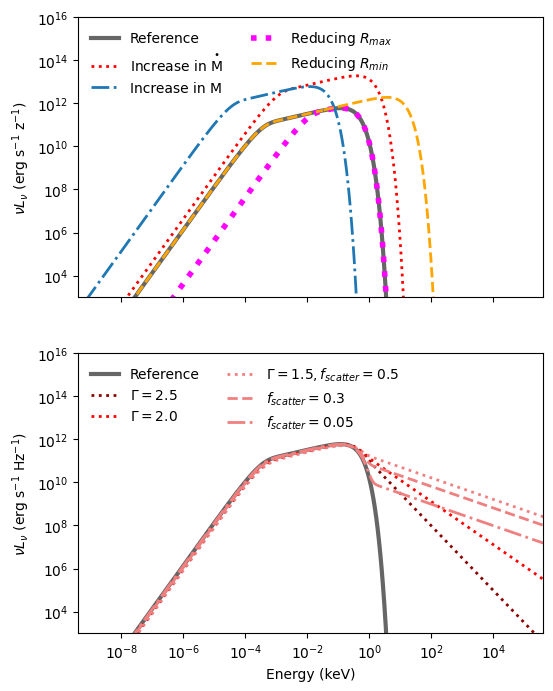}
    \caption{\textit{Top: }The SED for a thin disc which we use here to model X-ray emission. A reference disc SED is shown in gray for a 10 M$_\odot$ black hole accreting at 1\% of its Eddington accretion rate. The other curves show how changing a single parameter by two orders of magnitude affects the reference curve: increasing accretion rate $\dot{M}$ (dotted red curve), increase in black hole mass (dot-dashed blue curve), reducing maximum  disc radius (dotted magenta curve) and minimum disc radius (orange dashed curve). \textit{Bottom: }The figure shows distinct Comptonization curves for the same reference disc SED as in the top plot.  Dotted lines of different colours show different values for power law tail index ($\Gamma$) while distinct line styles show varying fractions of disc photons which get scattered ($f_{\rm scatter}$).}
    \label{fig:SED_parameter change}
\end{figure}

We illustrate how changes in parameters affect the disc SED in the top panel of Figure \ref{fig:SED_parameter change}. Increasing the mass of the accretor makes the spectrum softer, since as we increase the mass of the accretor the radius of the ISCO increases. On the other hand, increasing the accretion rate will make the XRB brighter and its spectrum harder. The inner and outer radii of the disc determine, respectively, the highest and lowest temperature a ring in the disc has and, thus, will dictate the range of frequencies in which the SED can be described by a superposition of black-body spectra.

\subsubsection{Fraction of Comptonized photons}

In addition to the disc spectrum, a high-energy tail component of emission is seen in virtually every SED of an X-ray binary and corresponds to the hard state of an XRB. As mentioned previously, these energetic photons are widely attributed to inverse-Compton scattering of soft photons by coronal electrons \citep{2006_Remillard}. 
The time spent by a binary on each one of the states (hard versus soft) varies from binary to binary.  Although it is known that transitions between states are related to changes in the accretion rate \citep{2013zhang, 2006Remillard}, they vary in time and strength of X-ray emission making it challenging to have a predictive model for when they should occur and how long they take \citep{2005Homan,Dunn2010, 2021Dong}. However, this change in X-ray production happens on short time scales of years or months. As we are concerned with the overall effect of X-rays on the environment on much longer time scales, the important quantity is the average amount of harder X-ray photons relative to the softer ones.  Thus, instead of modelling both states and trying to assign a time in which each would be active, we use observations to obtain a single SED for each binary which is a combination of its soft and hard states that correctly reproduces the ratio of soft to hard photons emitted during long stretches of time. In order to get an estimate for this ratio we use the publicly available data from the RXTE (Rossi X-ray Timing Explorer) survey\footnote{\label{fn: RXTE}\url{https://heasarc.gsfc.nasa.gov/docs/xte/archive.html}}. The RXTE data used for this paper spans over a decade of daily measurements of multiple sources out of which we selected 112 XRBs. RXTE is the survey with largest continuous observation of XRBs to which we have access. This allows us to reproduce an average XRB behaviour more reliably. It is important to point out that these XRBs are all from the local universe and, thus, are not representative of metal poor stars. Since we do not have long term data for any metal poor XRBs and given that we expect the mechanisms of X-ray emission to be alike, this is still a reasonable data set to be taken as a guideline for our models. The spectrum in X-rays is detected in 3 frequency ranges: $1.5-3$ keV (soft/S), $3-5$ keV (middle/M) and $5-12$ keV (hard/H). Note that the ranges of the RXTE soft and hard bands differ from the ones adopted in the rest of this paper, where we use $0.1-2$ keV as soft and  $2-10$ keV as hard for ease of comparison with the existing body of work on the high-redshift XRBs \citep[e.g.,][]{2013_Fragos, Fialkov_2014}.

\begin{figure}
    \centering
    \includegraphics[width = \columnwidth]{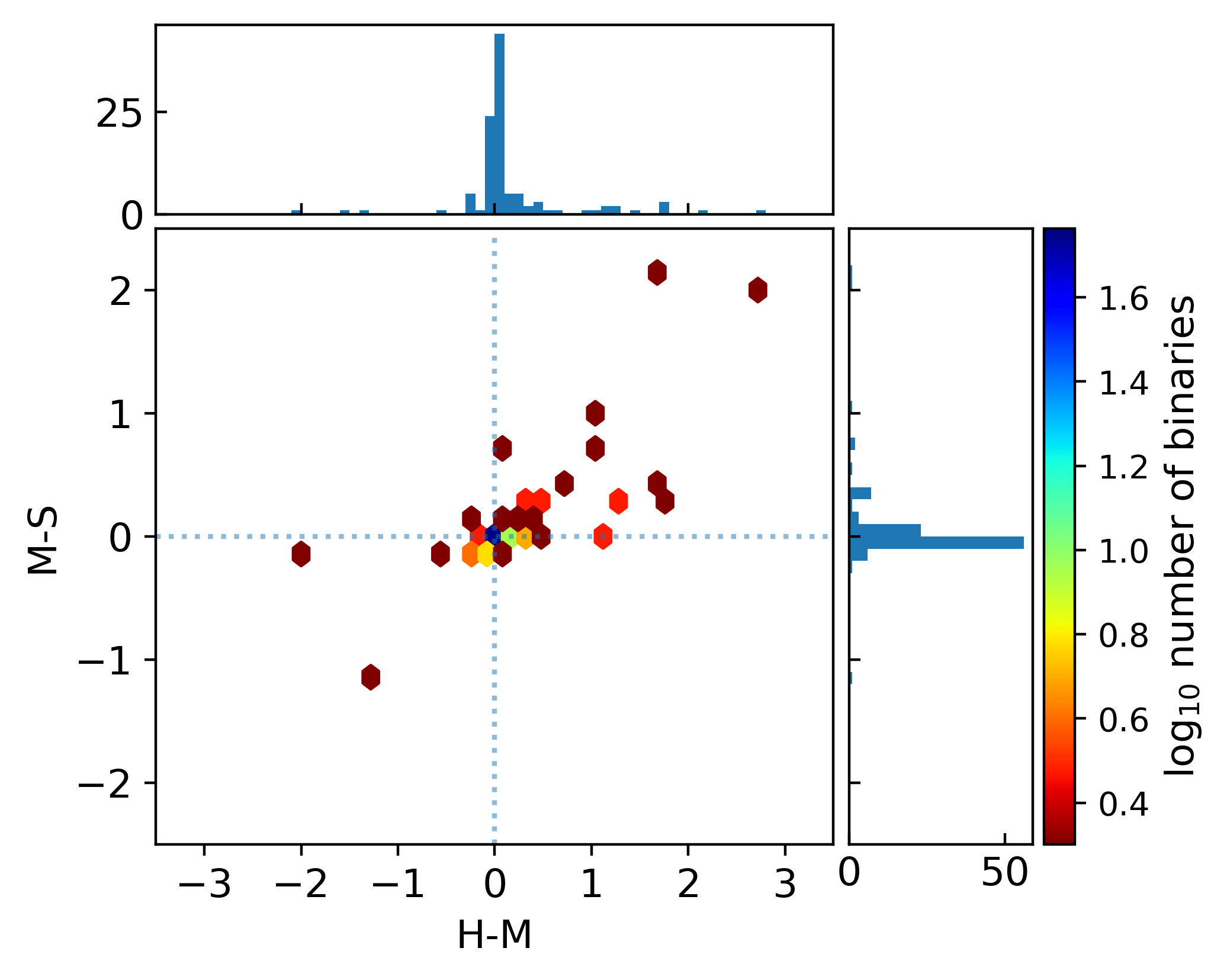}
    \caption{The RXTE counts of the middle band minus the soft band (y-axis) and the excess of the hard band counts over the middle band counts (x-axis). The counts are first summed in their respective bands for the length of time we have data available and then subtracted from one another. Each point on the plot represents an XRB. As can be seen, for most XRBs the number of photons emitted in the hard band are similar to the ones in the soft band (the sample is strongly clustered around the origin). It should be noted that, as the soft band has a smaller range of frequencies, the number of photons per frequency are still much larger for the softer end of the spectrum.}
    \label{fig: RXTE counts}
\end{figure}

In Figure \ref{fig: RXTE counts} we show a color-color diagram for all 112 binaries where we plot the excess number counts by the detector on RXTE of the middle band with respect to the soft band (M-S) versus the excess of the hard band (H-M) over the middle band. This gives an idea of the overall imbalance of hard versus soft photons. The majority of the observed  XRBs have a similar number of photons in the soft and hard bins as shown by data points being strongly clustered around the origin. Based on this observation, we adjust our Comptonization prescription in {\sc{binary\_c}} such that the final spectrum would lie within one standard deviation of the mean, that is the intersection of dashed lines in Figure \ref{fig: RXTE counts}.

\subsubsection{Comptonization spectrum}

The effect of thermal Comptonization on the photon distribution is described by the well-known Kompaneets partial differential equation \citep{Kompaneets_1957}. This equation, based on the Fokker–Planck formalism, describes the effect of multiple Compton scatterings on the photon distribution when the electrons are moving non-relativistically and the average fractional energy change per scattering is small. By solving the Kompaneets equation it can be demonstrated \citep{Rybicki_1986} that inverse Compton scattering will result in a power-law distribution of photon energies.  The solution for the up-scattered photons has a power law with index $\Gamma$.

In order to simulate the Compton up-scattering of the soft disc photons by the coronal electrons we use a convolution that converts a fraction of the disc photons to a Comptonized spectrum with a method analogous to the SIMPL model \citep{Steiner_2009}. 

Given an input distribution of photons $n_{\rm in}$ as a function of the initial photon energy $E_0$ and assuming a fraction $f_{\rm scatter}$ of the photons getting scattered, we can compute the output  distribution, $n_{\rm out}$, via
\begin{multline}
        n_{\rm out}(E)dE = \underbrace{(1 - f_{\text{ scatter}})n_{\text{in}}(E)\* dE}_\text{ unscattered photons}  \\ \qquad +\underbrace{f_{\text{ scatter}}\left[\int_{E_{\text{min}}}^{E_{\text{max}}} n_{\rm in}(E_0)G(E, E_0)dE_0 \right]\*dE}_\text{\rm scattered photons},
\end{multline}
where $G(E, E_0)$ is the Green's function describing the scattering. In the case of non-relativistic thermal electrons up-scattering ($E> E_0$):
\begin{equation}
    G(E, E_0)dE = (\Gamma -1)(E/E_0)^{-\Gamma} dE/E_0.        
\end{equation}
Thus, the effect of Compton scattering depends on two parameters $\Gamma$ and $f_{\rm scatter}$ as illustrated in the bottom plot of Fig. \ref{fig:SED_parameter change}.  $\Gamma$ regulates the slope of the tail whereas $f_{\rm scatter}$ determines to what point the original disc spectrum dominates over the Compton spectrum.

Most XRBs are observed to have an SED tail compatible with a value of $\Gamma$ between $1<\Gamma<3$ \citep{Yang_2015}. In our simulations we iterate to find a value for $\Gamma$ and $f_{\rm scatter}$ that gives approximately the value of hard to soft photons suggested by the RXTE data as is discussed in the previous section.

\subsection{Local absorption of X-rays}
\label{sec:absorption}

The luminosity and SED of XRBs discussed above (directly outputted by {\sc{binary\_c}}) are the intrinsic quantities produced by binaries. However, the radiation that reaches the IGM has different spectral properties owing to the soft X-rays being partially absorbed by the gas in the host XRB halo.  
Therefore,   in order to estimate the impact of X-ray emission on the IGM we need to take the local absorption of soft X-rays into account.  Here we provide a rough estimate of the optical depth at each X-ray frequency. It is beyond the scope of this paper to provide a more thorough analysis.

Given the gas density profile (our assumptions are outlined in Appendix \ref{app1}),  the optical depth $\tau$ for X-ray photons at frequency $\nu$ is 
\begin{equation}
    \tau(\nu) = \int_0^r (\sigma_{\rm H}(\nu) f_{\rm H}/m_{\rm H} + \sigma_{\text{He}}(\nu) f_{\text{He}}/m_{\rm He})\rho_{\rm gas}(r)dr,
\end{equation}
where $f_{\rm H} = 0.76$ and $f_{\text{He}}=0.25$ are the fractions of gas in atomic hydrogen and helium respectively by mass. The opacities for the first hydrogen and first helium ionization ($\sigma_{\rm H}$ and $\sigma_{\text{He}}$) are adopted from  \citet{Verner_1996}. For the halos considered here the average hydrogen column density is $\approx$ $10^{21}$ cm$^{-2}$. Using this optical depth, the fraction of photons that reach the IGM at a given  frequency $\nu$   is calculated $f_{\gamma}(\nu) = \exp\left(-\tau(\nu)\right)$. 
This formalism is applied in Section \ref{subsec: absorption} where we show the resultant XRB intrinsic and  absorbed spectra.



\section{Pop III systems that become X-ray binaries}
\label{sec: params that lead to XRBs}

Armed with our model, we now explore populations of high-redshift XRBs arising from the Pop III stars with different  IMFs (Table \ref{table: IMFs used}). A number of factors determine whether a binary becomes an XRB. In this Section we explore which of the initial orbital parameters and stellar masses are more likely to lead to the formation of an XRB. 

\subsection{Number of binaries}
\label{Sec: number of binaries}

\begin{figure}
    \centering
    \includegraphics[width = \columnwidth]{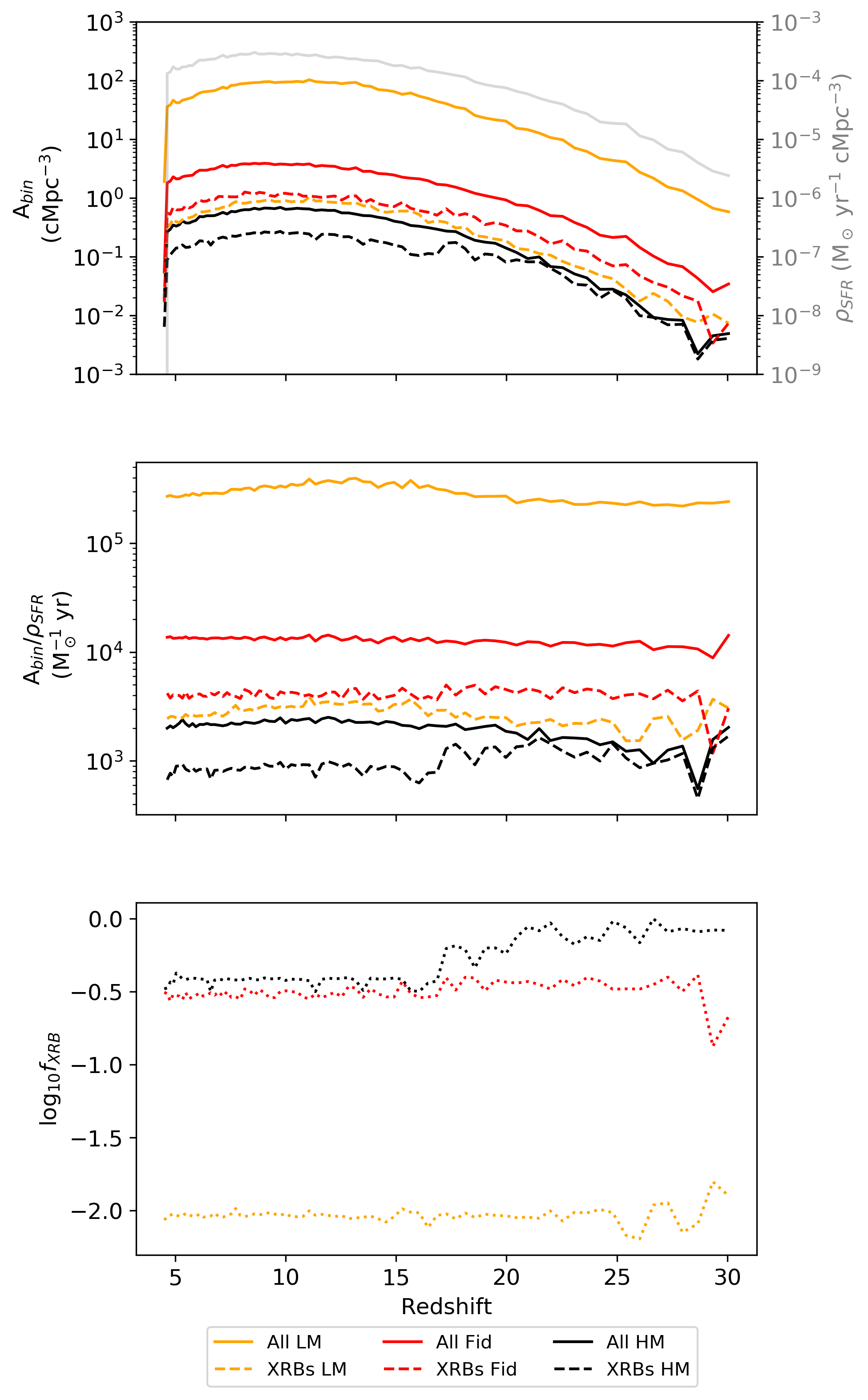}
    \caption{Evolution of the abundance ($A_{bin}$) of new binaries (or XRBs) with redshift. The Low-Mass, Fiducial and High-Mass IMFs are shown in yellow, red and brown respectively. \emph{Top panel:} Evolution of the total binary abundance per redshift. Solid lines represent all the binaries present in a catalogue, while dashed lines show only the binaries that evolve to become XRBs. The grey line and right scale show the SFR density for comparison. \emph{Mid panel:} Abundance of binaries per unit SFR density.  \emph{Bottom panel:} The dotted lines show the fraction of the binaries that result in  XRBs.  The fluctuations at  high-redshift present in all panels are due to the small number statistics and incomplete IMF sampling, as explained in the text. }
    \label{fig: number of XRBs}
\end{figure}

The most immediate factor that regulates the total number of XRBs per unit volume obtained with each IMF is the number of initial binaries present in each catalogue. Because the stellar mass in each dark matter halo at a given redshift is the same for all the IMFs, a more top-heavy IMF results in fewer stars than a bottom-heavy one. This leads to a large change in the total number of initial binary systems in each catalogue with the Low-Mass, Fiducial and High-Mass IMFs having initially $\sim 1\times 10^8, 4\times 10^7$ and $1\times 10^7$ binaries respectively (accounting for all binaries forming in the adopted 8 $(\rm{Mpc}/h)^3$ comoving box over all redshifts). The difference in the abundance of binaries between the catalogues is shown in the top panel of Figure \ref{fig: number of XRBs} where solid lines show the initial abundance of binaries at every redshift. 
The abundance of XRBs stemming from the binaries in each catalogue are shown in dashed lines. The abundance, ${A}_{\rm bin}$, of both initial binaries and XRBs grows with redshift as it follows the evolution of the Pop III star formation rate (SFR, shown in the top panel as a grey line).
We thus obtain a ratio of the binary abundance to the SFR density (${A}_{\rm bin}/\rm{\rho_{SFR}}$, middle panel of Figure \ref{fig: number of XRBs}) that is approximately constant. These values may be used in simulations in order to predict the number of XRBs and, thus, the ensuing X-ray background.

Owing to the dependence of binary-star evolution on the initial mass of the stars, a different fraction of binaries evolve into XRBs for different IMFs. We show this fraction $f_{\rm XRB}$ on the bottom panel of Figure \ref{fig: number of XRBs}. 

 This IMF-dependent XRB yield is a direct consequence of two factors, 
 (i) the abundance of binaries in each catalogue
  - the more binaries present at the start, the more XRBs may form - and (ii) the fraction of stars which are massive enough to give rise to a black hole or a neutron star necessary for the formation of an XRB.  For instance, the Low-Mass catalogue, in which a large fraction of the binary systems have both stars with masses below $8$ ${\rm M}_\odot$, only about 1\% of binaries form an XRB\footnote{Instead, such low-mass initial binaries live typically billions of years until they generate a white dwarf. Even if these systems lead to the formation of cataclysmic variables (a white dwarf accreting from a companion star), due to their long formation timescales, they would no longer play a role in the high-redshift X-ray emission.}, despite having the largest initial number of binaries.  In contrast, the High-Mass catalogue, which has fewer binary systems to start with, creates a substantial number of XRBs because of the larger number of massive stars present in this catalogue which reflects in its high $f_{\rm XRB}$ value. 
 
 The Fiducial IMF hits the sweet-spot of having an IMF with a substantial fraction of stars that is massive enough to lead to the formation of black holes and neutron stars, but not so massive that considerably limits the number of binaries able to form in a given halo. 
 
 We find that, averaging over redshifts $5-30$ (appropriately weighted by the time in each redshift), for a SFR density of $1 \, \mathrm{M}_\odot \, {\rm yr}^{-1}\, \rm{Mpc}^{-3}$ 2781, 4083 and 958 XRBs are present for the Low-Mass, Fiducial and the High-Mass catalogues respectively (Table \ref{tab:properties binaries}). For the fraction of binaries that become XRBs we have  $ f_{\rm XRB} = 9.3\times10^{-3}$, $3.2\times10^{-1}$ and $4.2\times10^{-1}$ for the  Low-Mass, Fiducial and High-Mass IMFs respectively. Here we see again that the Fiducial IMF is almost as efficient as the High-Mass IMF in producing XRBs.  

\begin{table}
    \centering
    \begin{tabular}{|c|c|c|c|}
    \hline &&&\\
         & Low-Mass & Fiducial &High-Mass  \\
    \hline 
   $M_{\rm{stellar}, {\rm min}}$  & 360 & 320 &1100\\
    $z_{\rm XRB}$ &27.6& 29.7 &  20.7\\
    $f_{\rm{XRB}}$ & $9.3\times10^{-3}$&$0.32$& $0.38  (0.42) $ \\
    $A_{\rm{XRB}}$/$\rho_{\rm SFR}$ & 2781 & 4083 & 958 (850)\\
    $L_{X, 0.1-2~{\rm keV}}/{\rho_{\rm SFR}}$ & $ 6.8\times 10^{38}$&$ 7.5\times10^{40}$&  $2.2\times10^{40}$ \\

    $L_{X, 2-10~{\rm keV}}/{\rho_{\rm SFR}}$  &$4.1\times10^{39}$&$ 4.3\times10^{41}$&  $7.9\times10^{40}$\\
    
    $L_{X, 0.1-2~{\rm keV}}/{\rm XRB}$ & $ 5.9\times10^{32}$&$ 3.3\times10^{34}$&  $5.8\times10^{34}$ \\
    
    $L_{X, 2-10~{\rm keV}}/{\rm XRB}$  &$ 2.9\times10^{33}$&$ 2.3\times10^{35}$&  $1.9\times10^{35}$\\

    $t_{\rm{XRB}}$ & $2.5\times10^{2}$ & 3.4& $3.7\times10^{-1}$\\
    $f_{\rm HMXB/LMXB}$ &  0.14 & 2.9& 12.2\\
    \hline
    \end{tabular}
    \caption{Properties for each catalogue. When averaging we take into account the number of binaries at redshifts $5<z<30$, unless otherwise specified, and weight it according to the time contained in each redshift bin. From top to bottom row: (i) the minimum stellar mass (${\rm M}_\odot$) in a halo in order for it to have at least one XRB ($M_{\odot}$) $M_{\rm{stellar}, {\rm min}}$; (ii) the redshift, $z_{\rm XRB}$, at which the first XRB appears in each catalogue; (iii) the average fraction over redshift, $f_{\rm{XRB}}$, of binaries that yield XRBs (for the High-Mass case we also give the average for $z<18$ in parenthesis which excludes the large fluctuations at high-redshift in which the High-Mass IMF is under-sampled), (iv) the abundance of XRBs per unit star formation rate, $A_{\rm{XRB}}$/SFR,  in units of ${\rm M}_\odot^{-1}$yr; (v)  luminosity per unit star formation rate in the soft X-ray band ($L_{X, 0.1-2~{\rm keV}}$, measured in $\rm{erg} s^{-1} {\rm M}_\odot^{-1} \rm{yr}$); (vi) luminosity per unit star formation rate in the hard X-ray band ($L_{X, 2-10~{\rm keV}}$, measured in $\rm{erg} s^{-1} {\rm M}_\odot^{-1} \rm{yr}$, $z<18$); 
    (vii/viii) luminosity per binary in the soft/hard X-ray band;
    (ix) average lifetime,  $t_{\rm{XRB}}$, of a XRB (in Myr) for all binaries at all redshifts; (x) the average fraction of HMXBs to LMXBs for each catalogue. 
}
    \label{tab:properties binaries}
\end{table}

At redshifts higher than $z = 20$ we clearly see strong fluctuations in all the quantities shown in Figure \ref{fig: number of XRBs} that particularly affect the number of XRBs. This stochasticity is due to the fact that the first star-forming haloes typically have a low stellar mass (just a few hundreds of solar masses) and, therefore,  contain a small number of binary stars which may, or may not, evolve into XRBs. The Low-Mass, Fiducial and High-Mass catalogues need to have, on average, at least 360, 320 and 1100 ${\rm M}_\odot$ in stars respectively to have at least one XRB (Table \ref{tab:properties binaries}). The Fiducial IMF requires halos to have the smallest mass for an XRB to form, indicating that this is the most efficient IMF when it comes to an early onset of X-ray feedback. The Low-Mass IMF requires a 12.5\% larger halo stellar mass than the Fiducial one because, as discussed above, most of the stars will have lower masses than what is needed for an XRB formation. Therefore, our Low-Mass catalogue, we need to sample more binary systems on average to find one with a binary that becomes a XRB. The High-Mass IMF has a different issue: owing to most of its stars being very massive, the halo stellar mass has to be 1100 is needed in order to form a binary system at all. That is  more than three times the one required by the Low-Mass and Fiducial IMFs,
However, if a binary system is created in the High-mass IMF it has a high likelihood of becoming an XRB as all binary pairs that are not in the PISN range potentially lead to an XRB.

Stemming from these differences between the catalogues, the history of XRB formation varies between the different IMFs. For example, the redshift at which there is at least one XRB in each halo, $z_{\rm XRB}$, is  27.6, 29.7 and  20.7 with the Low-Mass, Fiducial and High-Mass IMFs respectively (Table \ref{tab:properties binaries}). Even though we use a volume of 8 $(\rm{Mpc}/h)^3$, which should be sufficiently large the values quoted above are likely underestimated in all three models. XRB formation in the High-Mass case is delayed by $\sim 50$ Myr compared to the other two IMFs. This delay and the high characteristic stellar mass ( $M> 10 {\rm M}_\odot$ by construction) leads to stronger spatial fluctuations in the X-ray background created in the High-Mass IMF compared to the Low-Mass  and  Fiducial IMFs, which could have potential implications for observations of the CXB or the 21-cm signal.

It is not until $z\sim  22$ that the average halo stellar mass exceeds 2000 ${\rm M}_\odot$ and we are able to generate the full range of masses possible in the High-Mass IMF. Consequently, for the High-Mass catalogue the systems sampled at high-redshift tend to be biased towards lower mass stars ($10$ ${\rm M}_\odot < \,M/{\rm M}_\odot,\,< 180 $) and with a higher chance of becoming an XRB. This behaviour leads to the boosted binary fractions at high-redshifts. The reason why this range leads to a larger number of XRBs being formed is that it excludes stars that undergo PISN and also systems in which both stars are very massive,(${\rm M}> 260 {\rm M}_\odot$) which would photo-disintegrate.



\subsection{Masses of XRB binaries}
\label{sec: mass binaries}

The mass of the binary is decisive in whether or not the system becomes an XRB, and it is an important parameter that regulates its X-ray emission during the XRB stage. A binary needs to be massive enough for one of the stars to evolve into a black hole or a neutron star. On the other hand, binaries of stars which are initially too heavy could be impeded from becoming XRBs: stars in the PISN mass range ($\sim$ $180-260$ ${\rm M}_\odot$) leave no remnant behind after they explodes as a supernova. In addition, very massive ( $M> 260\,{\rm M}_\odot$) stars will encounter photo-disintegration instability. While these stars collapse to form a black hole, their lifetimes are so short that if two such stars are found in a binary they have no time to become an XRB. This is common in the High-mass scenario at smaller redshifts, when halos are massive enough to sample the whole IMF, and lead to the reduction in $f_{\rm XRB}$ from 0.42 at redshifts above 18 to 0.38 at later times (see Table \ref{tab:properties binaries}). 

\begin{figure}
    \centering
    \includegraphics[width=\linewidth]{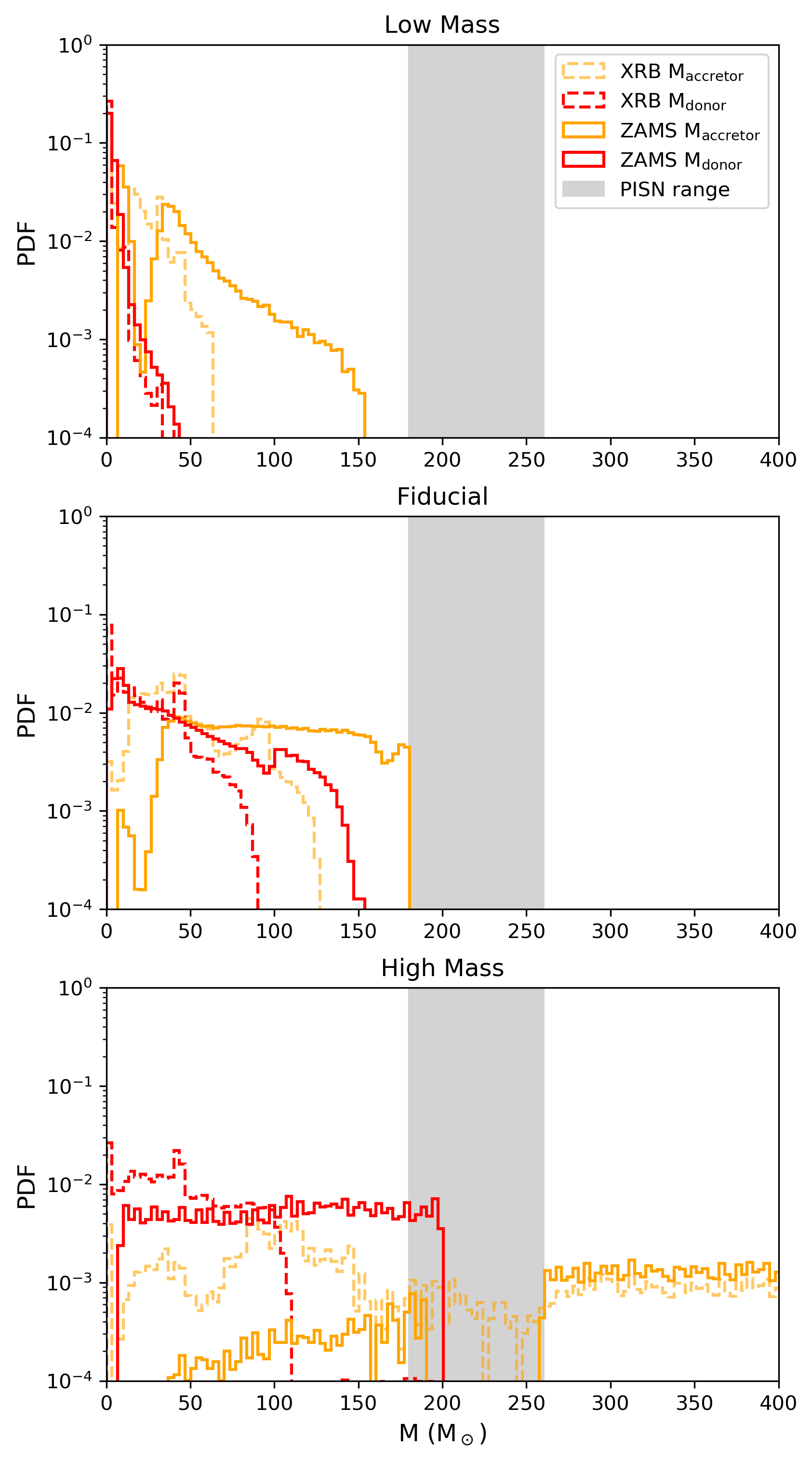}
    \caption{The probability density function of masses for stars in binaries that lead to the formation of XRBs both at the times when both stars are in the ZAMS (solid lines) and at the end of their time as XRBs  (dotted lines). The stars in the binary that were accretors are shown in orange and donors shown in red. The PISN range is shown by a grey rectangle. Each panel represents a different IMF (from top to bottom we have Low-Mass, Fiducial and High-Mass IMF)} 
    \label{fig:hist_masses}
\end{figure}

In Figure \ref{fig:hist_masses} we show the spread of masses of all binaries, at all redshifts, that passed through an XRB phase. We focus at two points in the binary lifetime: (i) at ZAMS (dashed lines) and (ii) at the end of X-rays emission (solid lines).
As expected, the accretor mass (yellow) is larger than the donor mass (red) at the ZAMS but, after the supernova, the remnant mass is significantly reduced, becoming about the same order or less than the mass of the secondary.
That is why the histogram of accretors at the end of the XRB stage is shifted to lower masses compared to the ZAMS.
The exception to this phenomenon are stars in excess of 260 M$_\odot$ (found only in the High-Mass case) which undergo photo-disintegration and result in a black hole of the almost the same mass as the star. 

Donor stars also lose a substantial part of their mass to the companion, which shifts their mass distribution below the ZAMS. In general the more massive donor the more mass they can lose both because they fill their Roche-lobe more easily, but also because they are massive enough for winds to play a role in their mass loss.  
The PISN region lies for masses between 180 and 260 $\rm{M}_\odot$ and is show by a grey rectangle. The Low-Mass and Fiducial cases have no XRBs in this range. The High-Mass case, however, has both accretors and donors within the PISN mass range. The remnants formed cover the entire mass gap, because black holes formed by PPISNe accrete and become larger (as to lie in the PISN mass range). Also, a small fraction of accretors have ZAMS masses at the PISN gap. These are stars that, before their death, manage to lose sufficient mass to die via PPISN and will lead to the formation of a black hole instead. 

\subsection{Orbital parameters}
\label{sec: orb params}
Orbital parameters of binaries are decisive in determining when mass can be transferred from the secondary star to the compact object. 
The main mechanism of mass transfer in our study is RLOF which occurs when the secondary star fills its Roche lobe and material flows to the remnant through the first Lagrange point. Because the exact location of $L_\mathrm{1}$ depends on the distance between the stars as well as on their masses, for any given remnant mass there are different orbital parameters that support the formation of an XRB. For instance, if the distance between the two objects in a binary is large, they might not interact at all and, thus, no X-rays are emitted. 
As laid out in Section \ref{sec: Methods}, we use observational distributions of orbital parameters suitable for present day main sequence binaries. In {\sc{binary\_c}} we trace the evolution of these parameters over time tracking changes resulting from mass exchange between the stars in each binary. 

\begin{figure*}
    \centering
    \includegraphics[width=\linewidth]{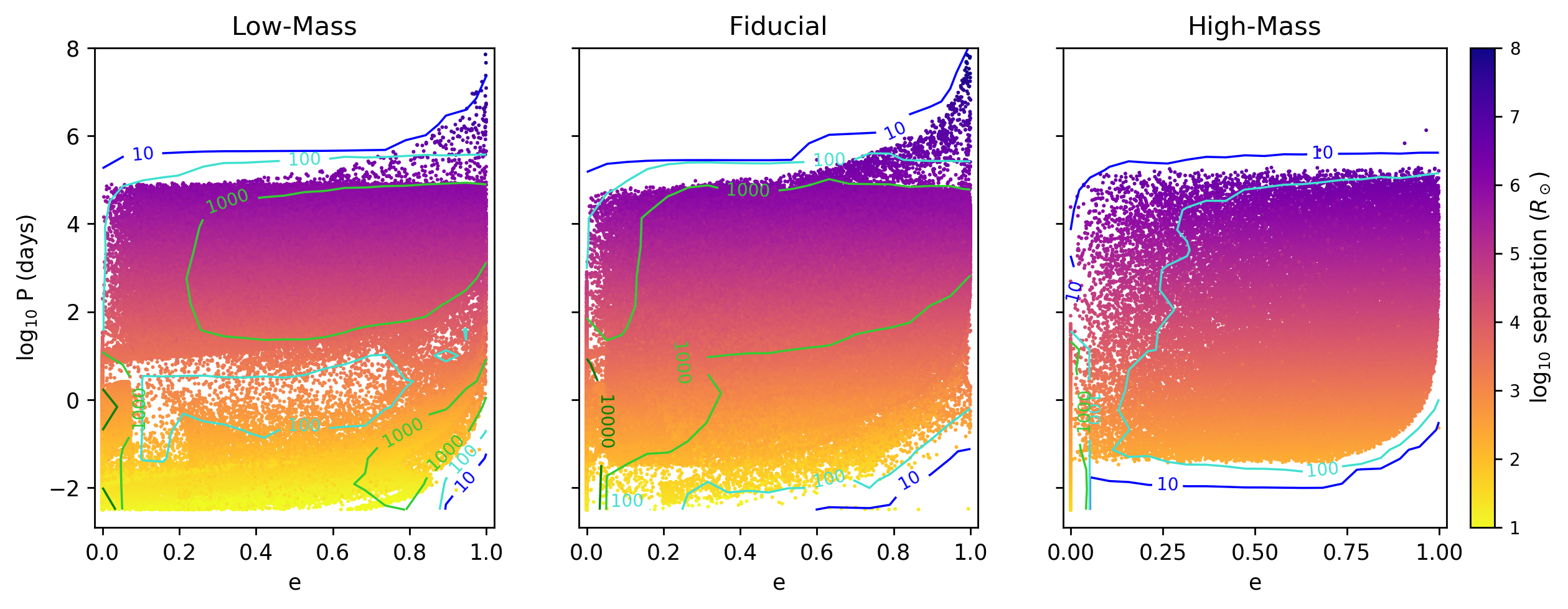}
    \caption{Logarithm of the period plotted against the eccentricity of the binaries that become XRBs in our simulation. From left to right we show the distributions for  the Low-Mass, Fiducial and High-Mass IMFs respectively. Each data point represents an XRB. The colour code indicates the semi-major axis of the stars in the binary in solar radii being, therefore, a measure of how far, on average, the stars are from each other while the contour lines show the number of systems in our simulation lying within a given region (contours shown for 10, 100, 1000, 10000 data points). In the cases of the Low-Mass and Fiducial  IMFs, systems with large average separations are able to harbour XRBs as long as the orbit of the binaries is highly eccentric. This can be seen by the trail of the points in the top-right corners of the left and middle panels. The High-Mass systems (right panel) do not exhibit such a behaviour and have, on average, smaller separations. The contour lines show that most systems are clustered in the left side of the plots, indicating that for all three IMFs the vast majority of systems that evolve into an XRB have been circularised (i.e. have close to zero eccentricity).}
    \label{fig:eccentricity}
\end{figure*}

In Figure \ref{fig:eccentricity} we show for each of our IMFs the period and eccentricity distributions the XRBs. Each point corresponds to an individual binary and is colour-coded to show the separation between the primary and the secondary.

We find that, in the cases of the Low-Mass and Fiducial catalogues (left and middle panels of Figure  \ref{fig:eccentricity}) a small number of long separation binaries evolve into XRBs. In these cases the orbit also has a large eccentricity. In other words, systems with large average separation still undergo mass transfer at perihelion and, thus, power X-ray emission. The highly-eccentric XRBs found in our simulations are remarkably similar to those observed which also have large eccentricities and separations and experience periodic accretion \citep{Belczynski_2009}, although these binaries are usually BeXRBs. 

Contrary to our result for the Low-Mass and Fiducial IMFs, we find no XRBs in systems with long separation or very long periods in the case of the High-Mass IMF (even though such systems are present in the initial catalogues). The lack of such systems is due to the short lifetimes of massive stars in the  High-Mass catalogue which are comparable to the long periods (of hundreds of thousands years). Therefore, even if such a system exists, it does not lead to an XRB.

While systems with long periods and high eccentricities do exist in our simulated Low-Mass and Fiducial catalogues, they are not the norm. This can be seen by the contour lines in Figure \ref{fig:eccentricity}. Indeed the vast majority ($>$99\%) of binaries are concentrated in circularised  systems. 

\section{X-ray luminosity of Pop III XRBs}
\label{sec:results}
In this Section we explore the X-ray luminosity of Pop III XRBs with different IMFs. We quote ratios of X-ray luminosity to SFR  in each scenario, which can be easily used in semi-analytical calculations and sub-grid models.




\subsection{Luminosity}

We consider the  X-ray luminosity in two bands, $0.1-2$ keV (soft band)  and $2-10$ keV (hard band). The average luminosity  is calculated by integrating the average XRB SED at a given redshift over the desired frequency range,

\begin{equation}
    L_{X,\nu_{\rm min}-\nu_{\rm max}} (z)/XRB = \int_{\nu_{\rm min}}^{\nu_{\rm max}} \frac{\left[\sum_{i=0}^N SED_i(z)\right]}{N} d\nu,
    \label{eq:12}
\end{equation}
where $N$ is the number of binaries at the given redshift $z$. 

\begin{figure}
    \centering
    \includegraphics[width=\columnwidth]{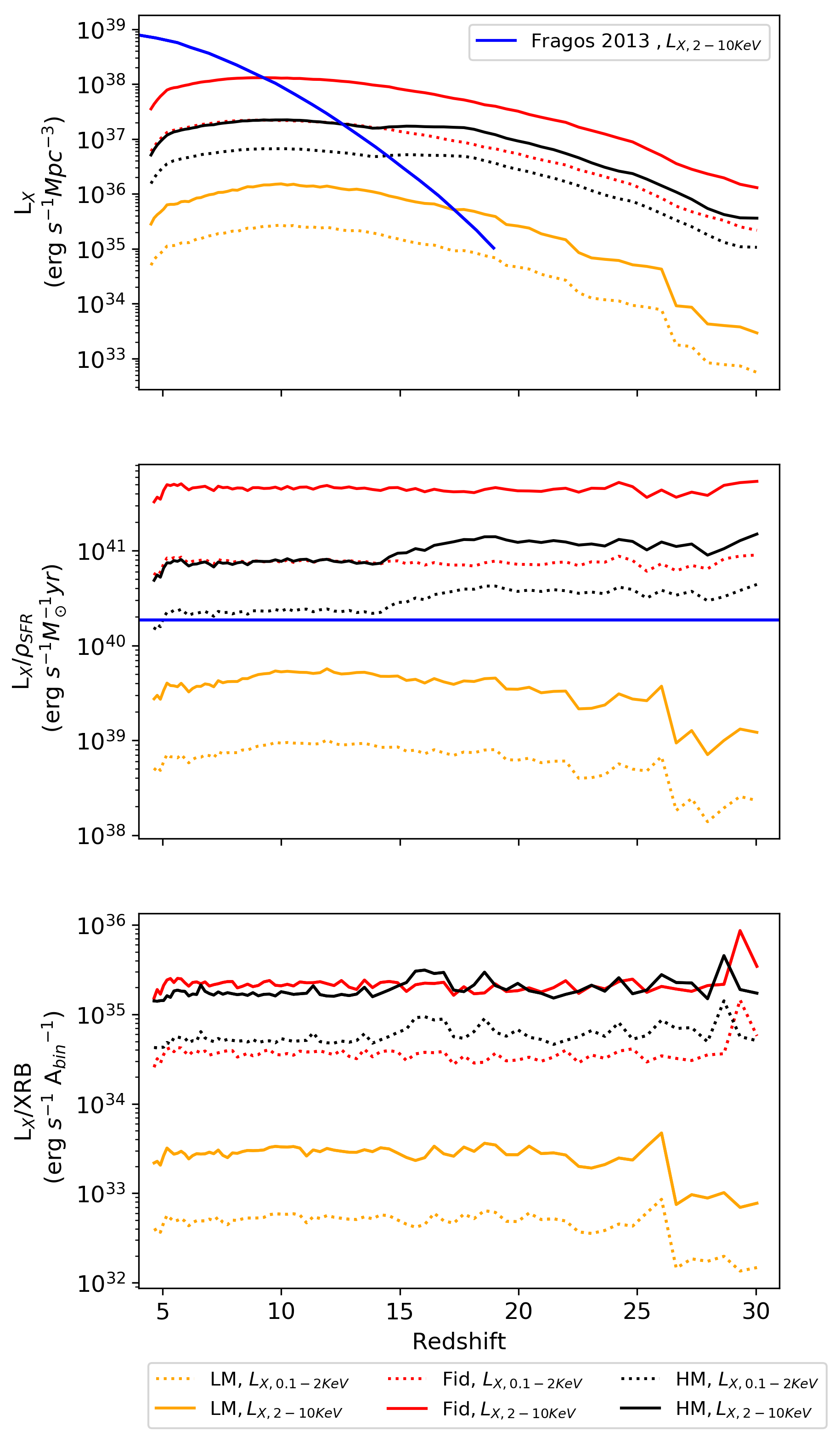}
    \caption{The average luminosity emitted in the X-rays for the soft ($0.1-2$ keV) and hard ($2-10$ keV) energy bands, shown in solid and dotted lines respectively. \textit{Top:} Redshift evolution of luminosity of an XRB population per comoving Mpc.  \textit{Middle} Luminosity per unit SFR. \textit{Bottom} The average luminosity emitted by a single binary. We show for comparison the luminosity expected for Pop II stars from \citet{Fragos_2013a} for the hard band in blue, assuming in for the curve in the middle panel a Pop II metallicity of Z=$10^{-4}$.}
    \label{fig:Luminosity_vs_z}
\end{figure}

The X-ray luminosity per unit volume is then  found by multiplying Eq. \ref{eq:12} by the XRB abundance computed in Section \ref{Sec: number of binaries}.
As mentioned previously, the soft band represents photons whose mean free paths are short enough to heat  and ionize  hydrogen atoms in the early Universe, be it in their original halo or in the IGM. Conversely, photons in the hard band have a mean free path which is longer than the horizon and, thus, would still be observed today as a part of the CXB (Section \ref{sec:CXB}). In Figure \ref{fig:Luminosity_vs_z}  we show the evolution of the luminosity in both bands with each of the IMFs considered. 

The top panel of Figure  \ref{fig:Luminosity_vs_z} shows the redshift evolution of luminosity density produced by Pop III XRBs. This includes all XRBs that started emitting X-rays at that point in time or that formed previously but are still in the XRB phase. The luminosity density is the highest, in both bands, for the Fiducial catalogue, followed by the High-Mass and the Low-Mass catalogues. The Fiducial and High-Mass catalogues contain brighter (more massive, see Figure \ref{fig:SED_parameter change}) XRBs compared to an average XRB in the Low-Mass case. In addition, the number of X-ray binaries is greater in the case of the Fiducial compared to the High-Mass IMF, which explains why the population with the Fiducial IMF is the most luminous. 

With all three IMFs the hard band (dotted lines) is more luminous than the soft band (solid lines). However, the SED for the XRBs peaks around $1-2$ keV  (Section \ref{subsec: LMXB and HMXB}) and that the higher luminosity of the hard band is in large part due to the band being wider than the soft band.

We can compare our results with the expected Pop II X-ray luminosity \citep{Fragos_2013a} represented by the blue curve.  Pop III X-ray luminosity dominates over Pop II at redshifts exceeding $z= 9.8, 13$ and $17$ with the Fiducial, High-Mass and Low-Mass IMFs respectively.

The middle panel of Figure \ref{fig:Luminosity_vs_z} shows the X-ray luminosity density per SFR. As the SFR is the same in all the three cases, the ratios between the different curves are the same as in the top panel.  We see that at low redshifts, $z<15$ for High-Mass and $z<20$ for the Fiducial case (i.e. past the initial period when the XRB population is poorly sampled), $L_{\rm X}$/SFR  settles to a constant value in both hard and soft bands (numbers are quoted in Table \ref{tab:properties binaries}).  In other words, for these two IMFs the X-ray luminosity produced by a population of the first XRBs is a good tracer of star formation. This owes to the fact that stars in the two catalogues are relatively massive and, thus, short lived. In contrast, as in Figure \ref{fig:Luminosity_vs_z}, the Low-Mass population is a poorer tracer of SFR ($L_X$/SFR varies even at low redshifts). Because the XRBs in the Low-Mass catalogue are formed from smaller stars, they take longer to form and also live longer lives (overall and as an XRB). As shown in Table \ref{tab:properties binaries}, XRBs from the Low-Mass catalogue live $2.4\times 10^{8} $ years which is almost three orders of magnitude longer than the XRB lifetimes in the Fiducial/High-Mass case. This results in a long delay between star formation and X-ray emission and in XRBs that formed at the highest redshifts still contributing to the X-ray emission at lower redshifts. 

The value of $L_X$/$\rho_{\rm SFR}$ in the hard band for low metallicity (Pop II) stars estimated in literature is of the order $\sim 10^{40} \rm{erg} \,\rm{s}^{-1}\,{\rm M}_\odot^{-1}\, yr$ \citep{2003_Glover,2013_Fragos,mineo_2013,Aird_2017} which is comparable to our Low-Mass catalogue. The more top-heavy IMFs (Fiducial and High-Mass) are expected to lead to a substantially higher X-ray emission per unit star formation density. The Fiducial a larger X-ray emission per unit star formation rate than the High-Mass IMF, despite their similar emissions per XRB (bottom panel of Figure \ref{fig:Luminosity_vs_z}). This is due to the larger number of XRBs formed per SFR in the Fiducial case (see Section \ref{Sec: number of binaries}). The High-Mass catalogue has the largest luminosity in the soft band per XRB due to the very high masses and accretion rates of these systems in comparison to the other IMFs. The hard band emission per binary is similar for both the Fiducial and the High-Mass catalogues, though due to the larger number of XRBs formed in the former the total emission in the hard band is higher than in the latter.
For the Low-Mass catalogue, which has not only has a smaller number of X-ray binaries, but also less massive accretors, the average luminosity per binary is lower by approximately two orders of magnitude compared to the other cases, despite the fact that a number of binaries formed at higher $z$ still contribute to the X-ray luminosity.



Our findings suggest that each IMF has a very distinct impact in the early Universe. 
\begin{itemize}
    \item At the dawn of star formation and in the case of the \textbf{High-Mass IMF}  XRBs will be rare (less than one per halo until redshift $z=20$) but bright ($L_X/XRB = 2.3\times10^{35}$ and $1.9\times10^{35}$ for the soft and hard band respectively).  Thus, we expect strong localised impact on the IGM and strong fluctuations in the CXB.
    \item Our \textbf{Fiducial IMF} leads to both abundant and efficient XRBs emitting both soft and hard X-rays. The expected  feedback on the IGM is more  uniform (compared to the case of the High-Mass IMF) and strong.
    \item Finally, in the case of the \textbf{Low-Mass IMF},  XRBs are present in almost every halo from the very start of star formation, but will be dimmer in X-rays. The feedback on the IGM is more uniform and weak.
\end{itemize}

\subsection{Cosmic X-ray background from Pop III XRBs}
\label{sec:CXB}

The cosmic X-ray background (CXB)  has been measured by deep X-ray surveys conducted using the Chandra X-ray Observatory \citep{Gilli_2007A&A...463...79G} and Swift \citep{Moretti_2012}. Although most of the CXB is known to be from resolved extra-Galactic X-ray sources, largely AGNs, a small amount of the CXB remains unresolved \citep{Moretti_2012, Gilli_2007A&A...463...79G}. This unresolved X-ray emission is able to set bounds on the
contribution of X-ray emission to reionization \citep{McQuinn_2012}. Using our data we can compute the contribution of Pop III stars to this measurement of the unresolved CXB.
X-rays emitted by Pop III stars that are detected today have redshifted and thus are now detected at a lower frequency. This effect plus the fact that soft X-rays interact with gas implies the contribution of soft photons emitted at high-redshift to the unconstrained CXB is negligible. We, therefore, use the hard band luminosity ($2-10$ keV) to estimate the contribution of Pop III X-ray binaries to the soft band ($0.5-2$ keV) of the CXB,
\begin{equation}
\begin{aligned}
    {\rm CXB}  = \frac{\Delta\Omega}{4 \pi}\frac{c}{H_0}\int_{z_{i}}^{z_{f}}\frac{L^{\rm V}_{ \rm X, 0.5(1+z)-2(1+z)~{\rm keV}}}{(1 + z)^2 \sqrt{\Omega_M (1+ z)^3 + \Omega_\Lambda}} \, dz \\ \quad \rm{erg \, s}^{-1} \rm{cm}^{-2} \rm{deg}^{-2},
    \label{eq:CXB}
    \end{aligned}
\end{equation}
where $L^V_{X}$ is the luminosity per unit volume observed in the $0.5-2$ keV band, $z_i =12$ and $z_f=30$ are the initial and final redshifts at which Pop III sources are expected to dominate \footnote{Population II binaries are likely to dominate X-ray emission at lower redshifts, while  at redshifts higher than 30 there are no sources in our catalogues.}. In Eq. \ref{eq:CXB}  $\Delta \Omega = 3.0 \times 10^{-4} \, \text{deg}^{-2}$ is the solid angle, $c$ is the speed of light, $H_0$ is the Hubble constant, $\Omega_M$ and $\Omega_\Lambda$ are the matter and dark energy density parameters respectively.

The contributions to the CXB ($0.5-2$ keV) of the Pop III XRB populations with different IMFs  are $6.52\times10^{-17}$, $9.17\times10^{-15}$ and  $2.58\times10^{-16}$ erg s$^{-1}$ cm$^{-2}$ deg$^{-2}$ with our Low, Fiducial and High-Mass IMFs respectively. All these contributions are small enough to be compatible with the observed unresolved CXB of $5^{+3.2}_{-2.6}\times10^{-12}$  erg s$^{-1}$ cm$^{-2}$ deg$^{-2}$ in the $2.0-10$ keV energy band \citep{Moretti_2012}. It is highly unlikely that Pop III XRBs contributed so significantly to the CXB, as the contribution to the unresolved CXB cited here ignores the X-ray production of Pop II stars \citep{Fragos_2013a} which, owing to the expected higher numbers of sources, would be a more important contributor thus saturating the observed CXB.   

\subsection{LMXBs and HMXBs}
\label{subsec: LMXB and HMXB}

As mentioned in Section \ref{sec: lm hm}, XRBs are often subdivided, according to the mass of their secondary stars, in LMXBs and HMXBs.  We set the dividing mass between the two categories to be 3 M$_\odot$ for ease of comparison with  \citet{Fragos_2013a}. 
Since the classification depends on the companion mass, it has a clear dependence on the IMF. Indeed, the ratio of the number of HMXBs to LMXBs ($f_{\rm HMXB/LMXB}$ in Table \ref{tab:properties binaries}) changes substantially for the IMFs we adopted, with the values for High-Mass, Fiducial and Low-Mass catalogues being 0.14, 2.9 and 12.1 respectively. As anticipated, the Low-Mass catalogue is the only one that has an overall dominance of LMXBs.

Because LMXB companions are less massive,  they live significantly  longer than HMXBs. Comparing the mean lifetime of XRBs with different IMFs, $t_{\rm{XRB}}$ (Table \ref{tab:properties binaries}), we find that binaries in the High-Mass catalogue emit X-rays for  $3.7\times10^{-1}$ Myr on average, while the mean lifetime is almost a thousand times longer ($2.5\times10^{2}$ Myr) in XRBs in the Low-Mass catalogue.
The short lifetimes of HMXBs makes them a suitable tracer of recent star formation  \citep{Grimm_2003, Mineo_2012}, while LMXBs are good estimators of the total stellar mass of a galaxy \citep{Gilfanov_2004}. We illustrate this behaviour in the middle panel of Figure \ref{fig:Luminosity_vs_z}, where it is clear that the luminosity of HMXB-dominated catalogues (i.e. the High-Mass and Fiducial)   follows the SFR significantly more closely than the Low-Mass case which is dominated by lower-mass binaries. 

\begin{figure}
    \includegraphics[width=\columnwidth]{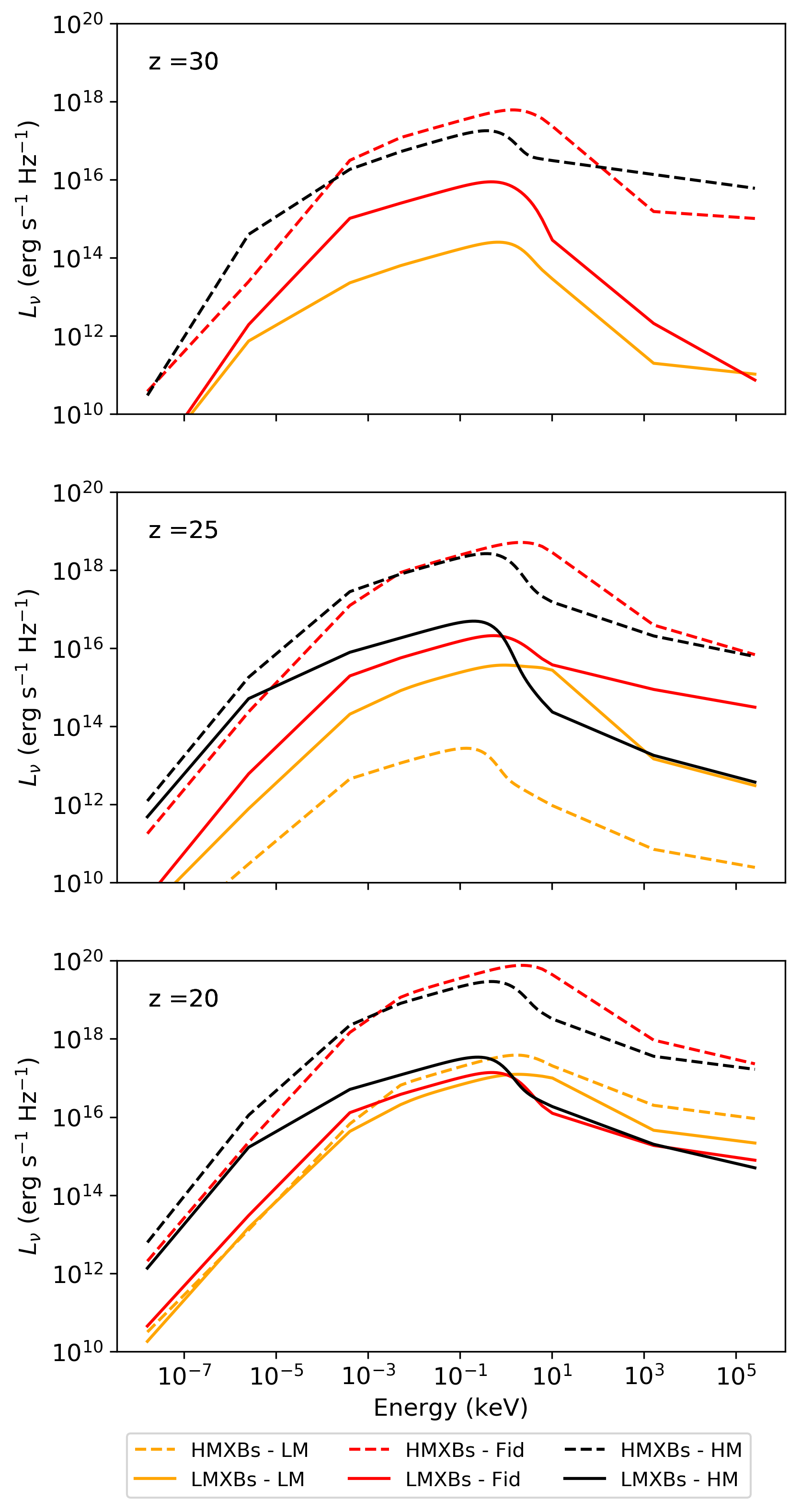}
  \caption{Comparison of the population-average SED of all LMXBs (solid) and HMXBs (dashed) at redshifts 20 (top), 25 (middle) and 30 (bottom) in our different catalogues. In general HMXBs are more luminous than LMXBs. Because the Low-Mass IMF is dominated by LMXBs, their emission dominates at high-redshifts when binaries with lower mass primaries are more common with the HMXBs beingnot present at redshift $z=30$. The High-Mass IMF initially ($z=30$) generates only HMXBs, since it only has stars with masses of 10 ${\rm M}_\odot$ or more. However, at lower redshifts ($z=25, 20$) we can see a very small fraction of these systems in which the companion star loses enough mass and is classified as a LMXBs.  }
  \label{fig: LMXB_HMXB}
\end{figure}

In Figure \ref{fig: LMXB_HMXB} we show the population-average SEDs of all LMXBs and HMXBs at redshifts $z=30$, 25 and 20 in each catalogue. 
At all redshifts the HMXBs of High-Mass catalogue dominate the emission at the lower frequencies, whereas the HMXBs of the Fiducial catalogue dominate the frequencies above 10 keV. At lower redshifts the contribution of HMXBs exceeds that of the LMXBs with all the considered IMFs in the entire X-ray range ($E > 0.1$ keV). Because HMXBs are the majority of the XRBs in the High-Mass and Fiducial catalogues, this reinforces the picture presented in the previous sections that the Fiducial IMF has the largest impact on the CXB and the High-Mass IMF having the strongest impact on the IGM. Interestingly, despite it only having stars with initial masses of 10 M$_\odot$ and above, we do find  LMXBs in  the High-Mass IMF case (although at relatively low abundance and not at the highest redshifts). This is due to the fact that some of the massive stars still manage to lose a substantial amount of mass either through interaction with their companion or in winds such that at their time as an XRB they will have a mass below 3${\rm M}_\odot$.

Only for our Low-Mass IMF is the LMXB SED is brighter than HMXB's. This only happens at $z=30$, however, as the sampling of this IMF is still biased towards lower masses. The trend reverts quickly with HMXBs and LMXBs having similar luminosities at $z=25$, and with the former becoming predominant at redshift 20 and lower.

Present day LMXBs and HMXBs often have different spectra due to their different accretion processes with the former also being typically less luminous. As mentioned previously in this study we consider only two accretion scenarios: RLOF and wind transfer. Although winds are present in HMXBs, they are still very weak. In addition winds are found in only a small fraction of the XRBs (less than 5\% even with the High-Mass IMF). As a result, SED shapes do not change significantly between LMXBs and HMXBs. The SED is instead more dependant on the average mass of the accretor, which tends to shift the peak of more massive IMFs towards smaller energies. Overall, whenever one excludes the possibility of BeXRBs, the differentiation between LMXBs and HMXBs is not vital in terms of determining the shape of the SED. While HMXBs are categorically  more luminous LMXBs, the ratio of their average luminosities ($L_{HMXB}/L_{LMXB}$) is strongly dependant on the companion mass chosen as the cutoff between the two categories. 


\subsection{Energy}
\label{sec: energy}

\begin{figure}
 \includegraphics[width =\columnwidth]{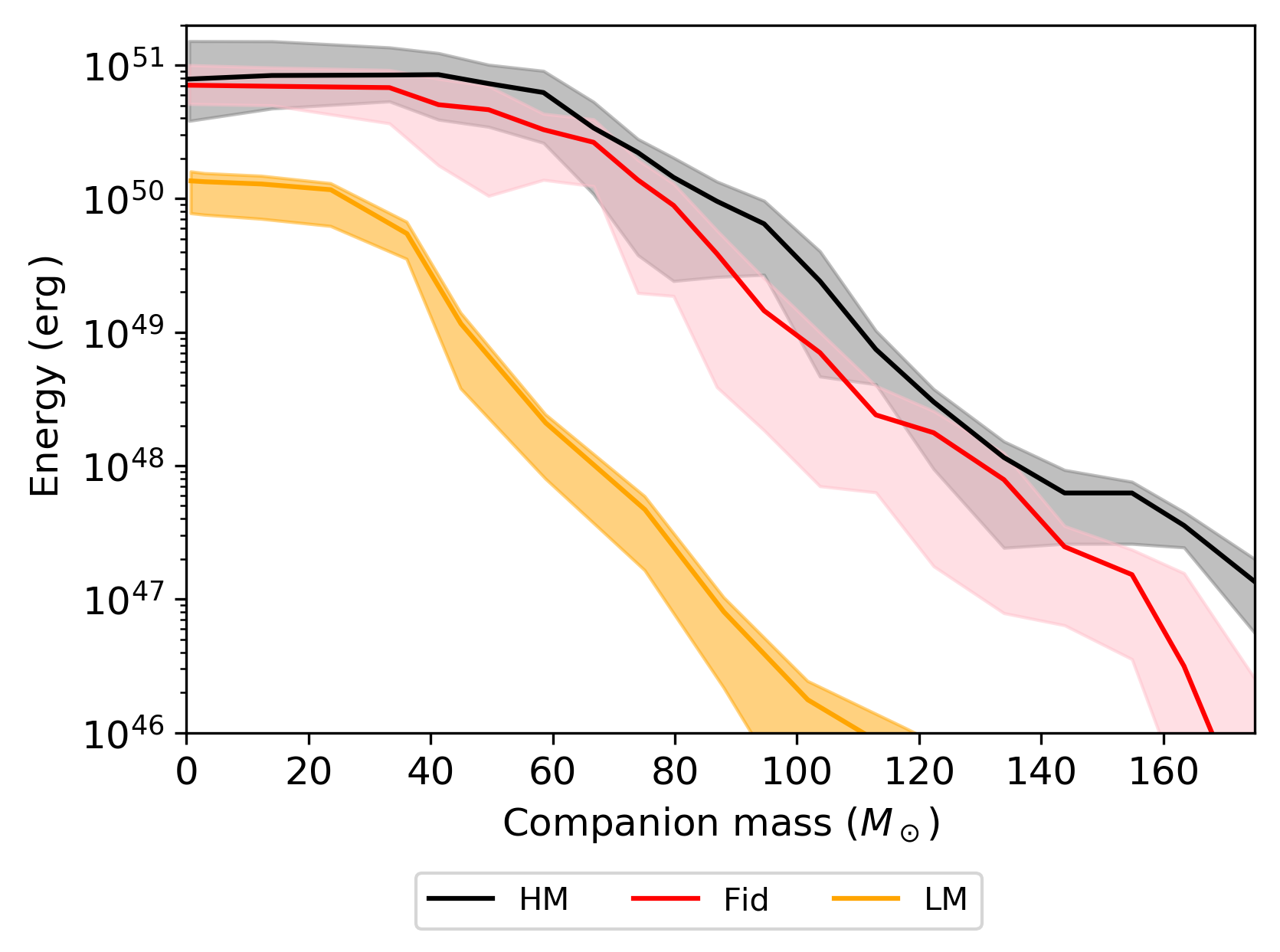}
  \caption{Mean total energy deposited during the lifetime of an XRB (binary emits X-rays) with a given companion mass for all three IMFs for all XRBs resulting from each catalogue. Lower and upper quartile shown as shaded regions. Despite having lower luminosities the binaries with a small companion mass live so much longer than their more massive counterparts that over their lifetime they output significantly more energy. We can see that this trend is applicable to all three IMFs.  This trend is also seen for the soft and hard bands separately, but we don't show them here for clarity}. 
    \label{fig: luminosity_time_and_mass}
\end{figure}

In Section \ref{subsec: LMXB and HMXB} we show that the luminosity of an XRB is higher the larger the companion mass due to increased accretion rates. Nevertheless, luminosity is not the only key factor when determining the impact on the IGM, but also the total energy deposited via X-ray emission. As some XRBs emit X-rays for much longer periods of time than others, looking solely at their instantaneous luminosity does not show the full picture of their impact in the IGM. To demonstrate this, we show in Figure \ref{fig: luminosity_time_and_mass} the average energy deposited by an XRB as a function of companion (donor) mass. In other words, we show the total X-ray luminosity (integrating the X-ray SED in the 0.5–100 keV range) multiplied by the time during which the binary emitted X-rays, averaged over XRBs with same companion mass. As it can be seen in  Figure \ref{fig: luminosity_time_and_mass}, lower mass companions, on average, deposit significantly more energy in X-rays over the course of their lifetime. This is due to the fact that lower mass companions tend to live longer as an XRB than their more massive counterparts. Thus, despite being dimmer (Figure \ref{fig: LMXB_HMXB}), low-mass  XRBs dominate the energy input owing to their much longer lifetimes and so their contribution cannot be ignored. The average energy deposited by a binary with a 3 M$_\odot$ companion is almost the same as by a star with a few tens of solar masses. However for companions in excess of 40 M$_\odot$ the energy deposition decreases rapidly, diminishing by more than two orders of magnitude for companions above $\sim$ 70/100/110 \, M$_\odot$ for Low-Mass, Fiducial and High-Mass IMFs respectively.

\subsection{Halo Absorption}
\label{subsec: absorption}
\begin{figure}
    \centering
    \includegraphics[width=\columnwidth]{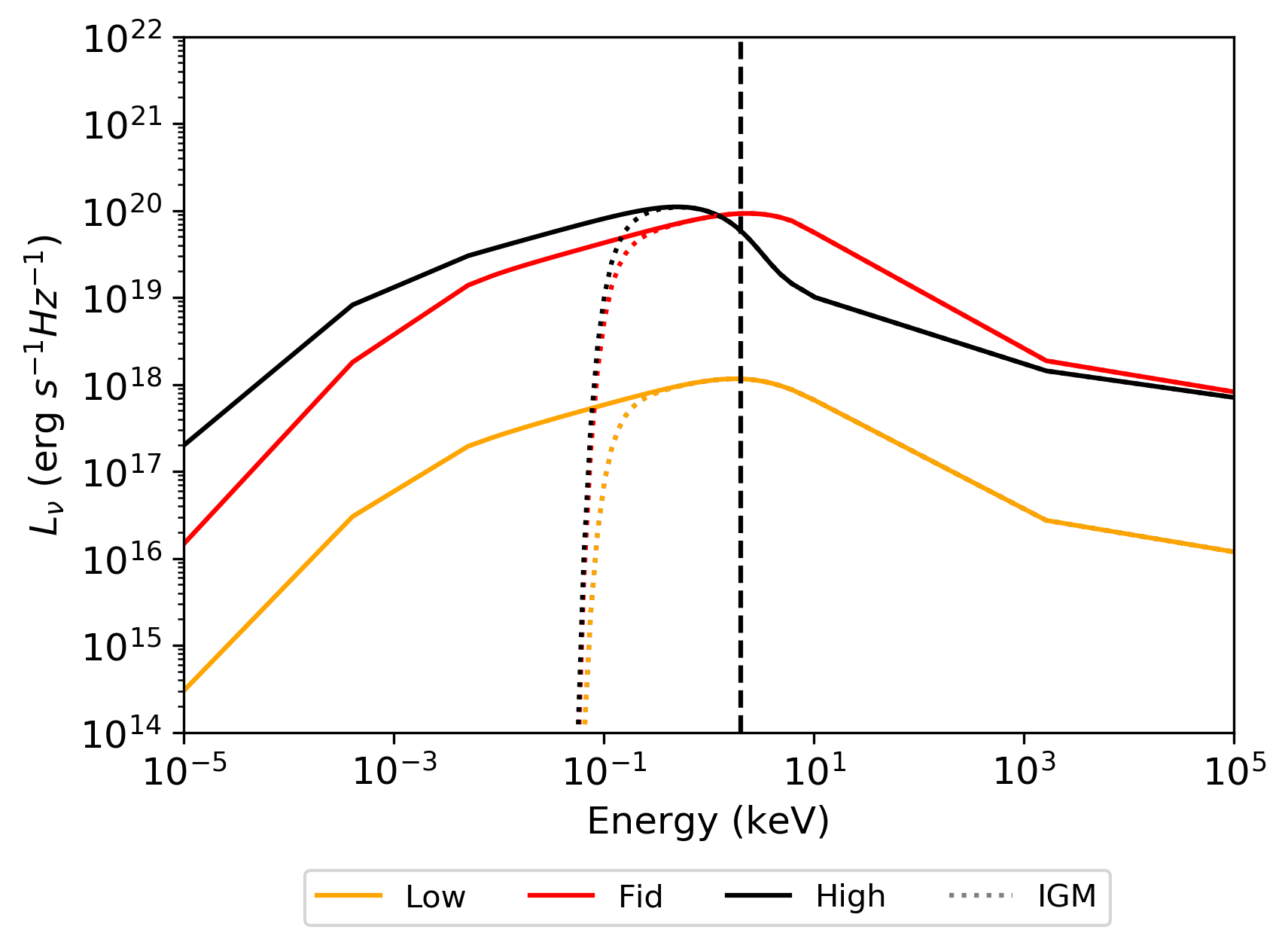}
    \caption{X-ray SEDs  averaged over the XRB population for each one of the considered IMFs considering absorption by the gas in the halo the XRB formed (dotted). We also show the un-absorbed emission in solid lines for comparison. The absorbed  SED is substantially harder than the intrinsic one. The black dashed line shows the rough division of the photons that will interact with IGM (to the left of the line) and the photons that will not (to the right). The exact location of the line depends on both the redshift and the neutral fraction of the IGM.}
    \label{fig:absorbed}
\end{figure}

Our results discussed above do not take into account the absorption of soft X-rays by gas present in the XRB host halo. Thus, our SEDs and luminosities over-predict the amount of soft X-ray emission that would reach the IGM. We apply the correction as discussed in Section \ref{sec:absorption} to estimate which photons reach the IGM. The comparison between the un-absorbed and the absorbed SEDs is shown in Figure \ref{fig:absorbed}. The division between the soft and the hard X-rays is indicated by a vertical dashed line. As we can see, soft photons with energies below 0.1 keV are completely absorbed and, thus, contribute nothing to IGM heating and ionization. Note that the contribution also depends on the redshift the X-ray emission takes place. The IGM is optically thick to photons with energy below $E$ \citep{2007pritchard}, where:
\begin{equation}
    E \sim 2 x_{\rm HI}^{1/3} \sqrt{ \frac{(1+z)}{15}} \, {\rm keV}
\end{equation}
where $x_{\rm HI}$ is the neutral fraction of hydrogen. Thus, assuming the IGM being neutral ($x_{\rm  HI}=1$) At $z=14$ all photons less energetic than 2keV are absorbed, at redshift $z=7$ only about 30\% is expected to still be neutral and only photons with energies below 0.43 keV experience an optically thick IGM.
Our estimate of the absorption optical depth is crude and the absorbed spectrum is shown here for illustration purposes only. More careful modelling of gas absorption of the host halo is required, but is beyond of the scope of this paper. 

\section{Conclusions}
\label{Sec:conclusions}


Understanding the formation and evolution of Pop III stars and their remnants, as well as quantifying their impact on the environment at high-redshifts, is becoming increasingly important with the growing interest in high-redshift observables such as the 21-cm signal of neutral hydrogen from cosmic dawn and bright prospects for sensitive observations of the X-ray sky at high-redshifts with telescopes such as ATHENA \citep{2013athena} and LYNX \citep{2019lynx}. 





We investigate how X-ray emission depends on the adopted IMF of Pop III stars. To bracket the uncertainty in Pop III properties, we considered three different IMFs: a Low-Mass (bottom-heavy IMF), a High-Mass (top-heavy) and a Fiducial one (and intermediate case), with the latter calibrated to reproduce late-time observables. We consider binaries fed by both stellar-winds and Roche-lobe overflow although, due to the low metallicity of Pop III stars atmospheres, wind accretion is sub-dominant. We explore statistical properties of XRBs in each case, including their number, orbital parameters at the X-ray phase and bolometric luminosity.

In Sections \ref{Sec: number of binaries}  and \ref{sec: mass binaries} we find that the abundance of XRBs is strongly dependant on the IMF chosen. This is due to two factors: (i) the initial number of binaries expected to form and (ii) the fraction of these systems that are candidates for XRB formation. The initial number of binary systems is proportional to the IMF slope as for a given stellar mass, such that more stars are formed for a bottom-heavy IMF. This is, however, counterbalanced by the fact that the more bottom-heavy the IMF, the lower is the number of stars massive enough to lead to the formation of black holes and neutron stars, and, thus, the smaller the number of systems that can potentially yield an XRB with less than a percent of the Low-Mass IMF catalogue binaries undergoing an XRB phase. Accordingly, the High-Mass IMF has a large fraction of binary systems that lead to the formation of an XRB, yet it does not lead to a large total number of XRBs due to the initially smaller number of binaries. As a result, we conclude that the optimal IMF to maximise the formation of XRBs is one with a moderate slope ($\sim -0.5$), such as the Fiducial IMF used here. 

Given the XRBs formed, we find that the High-Mass IMF results in more massive, brighter binaries than its Low-Mass counterpart. However, such massive XRBs have much shorter lifetimes. The trade-off between the higher luminosity and shorter lifetime regulates the total energy injected into the IGM (Section \ref{sec: energy}). We show that, combined, low-mass binaries deposit more energy in the environment over the course of their lifetime than binaries with more massive companions despite their lower luminosity. 

Considering the average SEDs of the XRBs produced, the High-Mass IMF tends to peak at lower energies than the other two IMFs. This implies that the top-heavy IMF has a larger number of photons that can heat and ionise either the IGM or the gas in the host halo. Conversely, a Low-Mass IMF leads to SEDs that peak at higher energies, having a harder spectrum overall. The Fiducial IMF has a similar energy peak for the SED as the Low-Mass one. Note, however, that due to the larger number of XRBs the Fiducial IMF would still have a larger number of soft photons reaching the IGM. 

Our results suggest that a High-Mass IMF leads to a stronger, but more spatially inhomogenous, X-ray emission in the early universe; whereas a Low-Mass IMF produces a weaker and more uniform X-ray feedback. The Fiducial IMF would produce an X-ray emission that is both strong and homogeneous. These differences are expected to have an effect on the 21-cm signal which strongly depends both on the intensity and the SED of X-ray sources \citep[e.g.][]{Fialkov_2014, Cohen2018}. Upper limits on the 21-cm signal produced by the existing radio telescopes \citep[HERA, LOFAR, SARAS,][]{HERAtheory, LOFAR1, Singh:2021} already constrain the IGM heating at $z\sim 6-10$. Although heating is a cumulative effect, and one needs to integrate over the entire cosmic history in order to derive the gas temperature at $z=6-10$, the contribution of Pop III heating at these redshifts is expected to be subdominant compared to that of Pop II.  More data from the existing telescopes, as well as the next generation 21-cm   experiments focusing on the cosmic dawn 21-cm signal, will shed light into the impact of Pop III XRBs on the IGM, thus allowing us to constrain properties such as the IMF.




\section*{Acknowledgements}

This work was performed using the Cambridge Service for Data Driven Discovery (CSD3), part of which is operated by the University of Cambridge Research Computing on behalf of the STFC DiRAC HPC Facility (www.dirac.ac.uk). The DiRAC component of CSD3 was funded by BEIS capital funding via STFC capital grants ST/P002307/1 and ST/R002452/1 and STFC operations grant ST/R00689X/1. DiRAC is part of the National e-Infrastructure. NS and AF were supported by the Royal Society University Research Fellowship. We would like to thank colleagues, Stefan Heimersheim, Thomas Guessey-Jones, Alex Tochter for the discussions we had over the content of this paper. TH acknowledges funding from JSPS KAKENHI Grant Numbers 19K23437 and 20K14464. RSK and SCOG acknowledge funding from the European Research Council via the ERC Synergy Grant ``ECOGAL -- Understanding our Galactic ecosystem: From the disk of the Milky Way to the formation sites of stars and planets'' (project ID 855130), from the Deutsche Forschungsgemeinschaft (DFG) via the Collaborative Research Center (SFB 881, Project-ID 138713538) ``The Milky Way System'' (sub-projects A1, B1, B2 and B8) and from the Heidelberg cluster of excellence (EXC 2181 - 390900948) ``STRUCTURES: A unifying approach to emergent phenomena in the physical world, mathematics, and complex data'', funded by the German Excellence Strategy. RGI and GM thank STFC for support (grants ST/R000603/1 and ST/L003910/1).

\section*{Data Availability}
The data underlying this article will be shared on reasonable request to the corresponding author.



\bibliographystyle{mnras}
\bibliography{ref.bib} 



\appendix

\section{Halo density profile}
\label{app1}

In this paper we calculate the density profile assuming a spherically symmetric halo of dark matter  mass $M_{\rm DM}$.  We also assume  the gas inside the halo to be  isothermal and in hydrostatic equilibrium, and  consisting  of atomic hydrogen (76\% of the gas) and helium (24\%). Assuming the gas has a negligible contribution to the total gravitational potential, we have
\begin{equation}
    \frac{kT_{\rm gas}}{\mu m_{\rm p}} \frac{\ln(\rho_{\rm gas}(r))}{dr} = - \frac{GM_{\rm DM}(r)}{r^2}
    \label{eq:hydroeq}
\end{equation}
where $\mu$ and $m_{\rm p}$ denote the mean molecular weight (which we adopt to be 1.32) and the proton mass. We adopt a  Navarro-Frank-White  profile \citep[NFW, ][]{Navarro_1996} for the dark matter, which  can be described fully by a concentration parameter  \citep[ $C=33$ $M_{\rm vir}$$/10^8{\rm M}_\odot$, ][]{Strigari_2007} and the halo mass, $M_{\rm  DM}(r)= M_{\rm vir} f(xC)/f(C)$,
where $f(a) = \ln(1+a) + a/(1/a)$ and $x=r/R_{\rm  vir}$ and $M_{\rm vir}$ and $R_{vir}$  are the virial mass and radius. Substituting the NFW profile into \ref{eq:hydroeq} we arrive to the expression for the density profile of a host halo, $\rho_{\rm  gas}(r)$
\begin{equation}
    \rho_{\rm gas}(r) = \exp\left[ \frac{G \mu m_{\rm p}}{kT_{\rm gas}}M_{\rm halo}\int_0^r\frac{f(xc)}{f(c)} dr\right].
    \label{eq:rho}
\end{equation}

 

\bsp	
\label{lastpage}
\end{document}